# The Design Principles of the Elixir Type System

Giuseppe Castagna[a], Guillaume Duboc[a,b], and José Valim[c]

a   IRIF, Université Paris Cité and CNRS, France
b   Remote Technology, France
c   Dashbit, Poland

**Abstract**   Elixir is a dynamically-typed functional language running on the Erlang Virtual Machine, designed for building scalable and maintainable applications. Its characteristics have earned it a surging adoption by hundreds of industrial actors and tens of thousands of developers. Static typing seems nowadays to be the most important request coming from the Elixir community. We present a gradual type system we plan to include in the Elixir compiler, outline its characteristics and design principles, and show by some short examples how to use it in practice.

Developing a static type system suitable for Erlang's family of languages has been an open research problem for almost two decades. Our system transposes to this family of languages a polymorphic type system with set-theoretic types and semantic subtyping. To do that, we had to improve and extend both semantic subtyping and the typing techniques thereof, to account for several characteristics of these languages—and of Elixir in particular—such as the arity of functions, the use of guards, a uniform treatment of records and dictionaries, the need for a new sound gradual typing discipline that does not rely on the insertion at compile time of specific run-time type-tests but, rather, takes into account both the type tests performed by the virtual machine and those explicitly added by the programmer.

The system presented here is "gradually" being implemented and integrated in Elixir, but a prototype implementation is already available.

The aim of this work is to serve as a longstanding reference that will be used to introduce types to Elixir programmers, as well as to hint at some future directions and possible evolutions of the Elixir language.

# The Art, Science, and Engineering of Programming

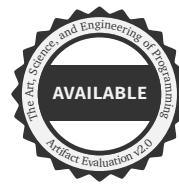
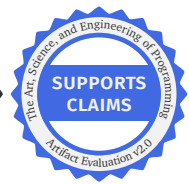

## 1 Introduction

Elixir [20] is a functional programming language that runs on BEAM, the Erlang Virtual Machine [48]. The language has been gaining adoption over the last years in areas such as web applications, embedded systems, data processing, and distributed systems, and used by companies like Discord and PepsiCo. The success of Elixir can be in good part ascribed to the underlying BEAM machine, developed by Ericsson in the eighties, and considered to be a great feat of engineering for concurrency, distribution, and fault tolerance. A limitation of both Erlang and Elixir is that they are dynamically typed, meaning they do not enjoy the safety features of a static type system that ensure at compile-time the absence of a given class of run-time errors.

Developing a static type system suitable for Erlang has been an open research problem for almost two decades. The earliest effort was attempted by Marlow and Wadler [34], which typed a subset of Erlang using subtyping unification constraints. However, their system was not adopted as type inference was slow, and the inferred types were large and complex. Ever since then, several attempts—either practical, theoretical, or both—have followed [37, 32, 50, 40, 21, 29, 45].

We present a gradual type system for Elixir, based on the framework of *semantic subtyping* [10, 23]. This framework, developed for and implemented by the CDuce programming language [3, 16], provides a type system centered on the use of set-theoretic types (unions, intersections, negations) that satisfy the commutativity and distributivity properties of the corresponding set-theoretic operations [23]. The system is a polymorphic type system with *local type inference*, that is, functions are explicitly annotated with types that may contain type variables, but their applications do not require explicit instantiations: the system deduces the right instantiations of every type variable. It also features precise typing of patterns and guards combined with *type narrowing*: the types of the capture variables of the pattern *and* of some variables of the matched expression are refined in the branches of case expressions to take into account the results of pattern matching. With respect to the system implemented for the CDuce language, the system we define for Elixir brings several novelties and new features. Its main contributions can be summarized as follows:

- *Semantic subtyping.* An extension of the semantic subtyping framework to fit Elixir/Erlang, in particular, the definition of new function domains to account for the tight connection between Elixir/Erlang functions and their arity.
- *Guards.* A new typing technique for analyzing guards in pattern matching.
- *Records and dictionaries.* A new typing discipline unifying records and dictionaries.
- *Dynamic type.* The integration of the dynamic type in the type system, which is used to describe untyped parts of the code, and how they interact with statically typed parts. This uses and innovates techniques of the gradual typing literature.
- *Strong arrows.* A new gradual typing technique for typing functions that takes into account runtime type tests performed by the virtual machine or inserted by the programmer. This makes it possible to guarantee the soundness of the gradual typing system with precise static types, without modifying the compilation of the source code, that is, in the terminology of [31], a *safe* type system with *erasure gradual typing*.

The system we present is, thus, a combination of both existing (set-theoretic types, polymorphism with local type inference, narrowing, gradual typing) and new (guard analysis, strong arrows, records and dictionaries) techniques. We believe that for the definition of a type system for the Elixir/Erlang ecosystem[1] there is no single silver bullet solution, and that, instead, *all* the techniques above are necessary to give the system a chance to be accepted and adopted by the community of developers (see also Section 5).

We partially implemented the system by a simple prototype that was internally tested by the Elixir development team and that received scattered feedback from the Elixir community. At the moment of writing we are working on an implementation to (gradually) integrate the system in the compiler that we hope to test in the course of next year. Only then we will be able to assess the usability of the type system, how it will perform on large codebases, and the degree of adoption by the community. And while we are cautiously confident in the work we present here, we do not rule out the possibility of finding unforeseen deal-breakers that could send us back to square one.

**Outline**   In Section 2, we provide an overview of the specific typing issues that arise in Elixir and the reasons why set-theoretic types are a good fit to type it. In Section 3, we demonstrate the various typing techniques we developed specifically for Elixir. We outline the formal approach to typing programs in Elixir in Section 4. Section 5 covers our design principles for integrating the type system into the Elixir compiler and the impact on programmers. Section 6 discusses related work, while Section 7 concludes our presentation and outlines some features planned as future work.

## 2   Typing Elixir: An Overview

### 2.1   Why Types

Static typing seems nowadays to be the biggest need for the Elixir community. Today, Elixir supports Typespec, a mechanism for annotating functions with Types [49]. However, Typespec's specifications are not verified by the compiler. Instead, a tool called Dialyzer, which ships with the Erlang standard library, can be used to find discrepancies in your source code and type annotations. Dialyzer is based on success typing [32], which guarantees no false positives, but may leave several bugs uncaught.

As the Elixir community grows, the general feedback is that, while Dialyzer is helpful and provides developers with some guarantees, its ergonomics and functionality do not fully match the community expectations. Based on our experience with the language and its ecosystem, we speculate developers would accept more false positives from the compiler in exchange for catching more bugs. Hence, the interest of the authors in fully baking static typing into the Elixir compiler.

---

[1] Semantically, Elixir is a superset of Erlang with the addition of protocols and macros. This is a design goal of Elixir, reflected by the fact the Elixir front-end compiler produces Erlang AST. Therefore, by typing all Elixir idioms, we also type all Erlang idioms.

The benefits we expect are essentially twofold. The first benefit of types is to *aid* documentation (emphasis on the word "aid" since we don't believe types can replace textual documentation). Elixir already reaps similar benefits from Typespec, and we expect an integrated type system to be even more valuable in this area.

The second benefit of static types revolves around contracts. If function `caller(arg)` calls a function named `callee(arg)`, we want to guarantee that, as these functions change over time, the caller passes valid arguments into the callee and correctly handles the return types from the callee.

This may seem like a simple guarantee, but we can run into tricky scenarios even on small code samples. For example, imagine that we define a `negate` function, that negates numbers. One may implement it like this:

```
1 $ integer() -> integer()
2 def negate(x) when is_integer(x), do: -x
```

The `negate` function receives an `integer()` and returns an `integer()`.[2] Type specifications are prefixed by $ and each specification applies to the definition it precedes. With our custom negation in hand, we can implement a custom subtraction:

```
3 $ (integer(), integer()) -> integer()
4 def subtract(a, b) when is_integer(a) and is_integer(b) do
5    a + negate(b)
6 end
```

This would all work and typecheck as expected, as we are only working with integers.

Now, imagine in the future someone decides to make `negate` polymorphic (here, *ad hoc* polymorphic), by including an additional clause so it also negates booleans:

```
7 $ (integer() or boolean()) -> (integer() or boolean())
8 def negate(x) when is_integer(x), do: -x
9 def negate(x) when is_boolean(x), do: not x
```

The specification at issue uses `integer() or boolean()` stating that both the argument and the result are either an integer or a boolean. This is a union type which has become common place in many programming languages.

The type specified for `negate` is not precise enough for the type system to deduce that when `negate` is applied to an integer the result is also an integer.

```
10 Type warning:
11    | def subtract(a, b) when is_integer(a) and is_integer(b) do
12    |   a + negate(b)
13        ^ the operator + expects integer(), integer() as arguments,
14          but the second argument can be integer() or boolean()
```

Such a type system would not be enough to capture many of Elixir idioms, and it would probably lead to too many false positives. Therefore, in order to evolve contracts over time, we need more expressive types. In particular, to solve this issue we need an *intersection type*, which specifies that `negate` has both type `integer()->integer()` (i.e., it is a function that maps integers to integers) *and* type `boolean()->boolean()` (i.e., it is a function that maps booleans to booleans). This type is more precise than the previous one and is written as:

```
15  $ (integer() -> integer()) and (boolean() -> boolean())
```

---

[2] We follow Erlang convention that basic types are suffixed by "`()`", for instance, `string()`.

With this type, the type checker can infer that applying `negate` to an integer will return an integer. Therefore, in the definition of `subtract`, the application `negate(b)` has type `integer()`, and the function `subtract` is well-typed.

## 2.2 Set-Theoretic Types and Subtyping Relation

Unions, intersections, and—see later on—negations are called *set-theoretic* types, insofar as they can be thought of in terms of sets: if we think of a type as the set of all values of that type (e.g., `integers()` as the set of all integer constants, `boolean()` as the set containing just `true` and `false`, ...), then the union of two types is the set that contains the union of their values (e.g., a value of type `integer()` `or` `boolean()` is either an integer value or a boolean value), the intersection of two types is the set that contains the values that are in both types (e.g, a value in the intersection `(integer()->integer())` `and` `(boolean()->integer())` is a function that both maps integers to integers and maps booleans to integers), and, finally, the negation of a type is its complement, that is, it contains all the (well-typed) values that are not in the type (e.g., a value in `not` `integer()` is any value that is not an integer).

Notice that an intersection of arrows does not necessarily correspond to multiple definitions of a function. For instance, the following definition is well-typed:

```
16  $ (integer() -> integer()) and (boolean() -> boolean())
17  def negate_alt(x), do: (if is_integer(x), do: -x, else: not x)
```

We have seen that we can specify two different types for `negate`, that is:

(1) `(integer() or boolean()) -> (integer() or boolean())`
(2) `(integer() -> integer()) and (boolean() -> boolean())`

and we said that the latter type is "more precise" than the former. Formally, we state that the latter is a *subtype* of the former, meaning that every value of the latter is also a value of the former. In the case of the two types above, the subtyping relation is also *strict*: every function that maps integers to integers and booleans to booleans, is also a function that maps an integer or a boolean to an integer or a boolean, but not vice versa. For example, the constant function `fn x -> 42 end` maps both integers and booleans to integers and thus to `integer()` `or` `boolean()`; as such it is a function of the type in (1). However, it does not map booleans to booleans. Therefore, it is not in the intersection type in (2). When two types are one subtype of each other they are said to be *equivalent*, since they denote the same set of values (e.g., `(integer()->integer())` `and` `(boolean()->integer())` is equivalent to `(integer` `or` `boolean())->integer()`).

The type of `negate` or `negate_alt` can also be expressed without intersections, by using parametric bounded quantification,[3] but this is seldom the case. For instance,

---

[3] Precisely as `$ a -> a` `when` `a: integer()` `or` `boolean()` : see Section 2.3.

Elixir provides a negation operator named `!`, which is defined for all values. The values `nil` and `false` return `true`, while all other values return `false`. With set-theoretic types, we can give to this operator the following intersection type:

```
18 $ (false or nil -> true) and (not (false or nil) -> false)
```

This type introduces two further ingredients of our type syntax: singleton types and negation types. Namely, the atoms `true`, `false`, and `nil`,[4] are also types, called singleton types, because they contain only the constant/atom of the same name. The connective **not** denotes the negation of a type, that is, the type that contains all the well-typed values that are not in the negated type, whence the interpretation of the functional type above.[5]

The advantage of interpreting types as the set of their values is that types satisfy the distributivity and commutativity laws of their set-theoretic counterparts.

For instance, a well-known property of products is that unions of products with a same projection factorize, that is, `{s1,t} or {s2,t}` is equivalent to `{s1 or s2, t}` (Elixir uses curly brackets for products). This is reflected by the behavior of our type-checker that accepts the following definitions:

```
19 $ type t() = {integer() or string(), boolean()}
20 $ type s() = {integer(), boolean()} or {string(), boolean()}
21 $ ((t() -> t()), s()) -> s()
22 def apply(f,x) do: f.(x)
```

The first two lines define the types `t()` and `s()` while lines 21-22 define a function whose typing demonstrates that the type-checker considers `t()` and `s()` to be equivalent. This is because it allows an expression of type `t()` to be used where an expression of type `s()` is expected (i.e., `f` which expects an argument of type `t()` is given an argument `x`, which is of type `s()`) and an expression of type `s()` where an expression of type `t()` is expected (i.e., the type specification declares that `apply` returns a result of type `s()`, but the body returns `f(x)` which is of type `t()`). In contrast, languages that use a syntactic definition of subtyping, such as Typed Racket, Flow, or TypeScript, accept the application `f(x)` but reject the typing of `apply`: they cannot deduce that `t()` is a subtype of `s()`.

Finally, we adopt the Typespec conventions wherein `term()` represents the top type (i.e., the type of *all* values) and `none()` denotes the *empty* type, that is, the type that has no value and which is equivalent to **not** `term()` (likewise, `term()` is equivalent to **not** `none()`).

### 2.3 Applying Set-Theoretic Types to Elixir

The existing set-theoretic types literature enables our type system to represent several Elixir idioms. We outline some examples in this section.

---

[4] In Elixir, atoms are user-defined constants obtained by prefixing an identifier by colon, as in `:ok`, `:error`, and so on. The atoms `true`, `false`, and `nil` are supported without colon for convenience.

[5] The precedence of **and** and **or** is higher than type constructors (arrows, tuples, records, lists), and the negation **not** has the highest precedence of them all.

**Parametric Polymorphism with Local Type Inference**    Set-theoretic type-systems feature parametric polymorphism with local type inference: expressions (in particular functions) can be given types containing type variables, but to use them it is not necessary to specify how to instantiate these variables, since the system deduces it [15, 14].

In our implementation, type variables are identifiers that are quantified by using a postfix **when** in which variables come with their upper bound.[6] Type variables are distinguishable from basic types, since they are *not* suffixed by "`()`". We feature only first order polymorphism, so **when** can only occur outside a type (never inside it).

The `map` and `reduce` operations over lists are good examples of need for polymorphic types, since most of the functions working with collections (known as "enumerables" in Elixir) cannot be sensibly typed without them. For instance, we have

```
23  $ ([a], (a -> b)) -> [b] when a: term(), b: term()
24  def map([h | t], fun), do: [fun.(h) | map(t, fun)]
25  def map([], _fun), do: []
```

```
26  $ ([a], b, (a, b -> b)) -> b when a: term(), b: term()
27  def reduce([h | t], acc, fun), do: reduce(t, fun.(h, acc), fun)
28  def reduce([], acc, _fun), do: acc
```

meaning that for all types `a` and `b` (i.e., for all `a` and `b` subtypes of `term()`):

- `map` is a binary function that takes a list of elements of type `a` (notation `[a]`), a function from `a` to `b` and returns a list of elements of type `b`;
- `reduce` is a ternary function that takes a list of `a` elements, an initial value of type `b`, a binary function that maps `a`'s and `b`'s into `b`'s, and returns a `b` result.

Local type inference infers that for `map([1, 4]`, **fn** `x -> negate(x)` **end**) both type variables must be instantiated by `integer()`, deducing the type `[integer()]` for it.

Intersection can also be used to define the type specification of `reduce` for the case of empty lists (in which case the third argument can be of any type):

```
29  $ (([a] and not [], b, (a, b -> b)) -> b) and
30    (([], b, term()) -> b) when a: term(), b: term()
```

Polymorphic types make inference more precise for other functions. For instance, if we add a default case to the `negate` example (lines 51-52) we obtain the code

```
31  def negate(x) when is_integer(x), do: -x
32  def negate(x) when is_boolean(x), do: not x
33  def negate(x), do: x
```

for which we can deduce—or at least check—the type (notice the use of bounded quantification in line 36)

---

[6] We did not specify lower bounds since they are not frequently used and they can be encoded by union types, e.g., $\forall (s \leq \alpha).\alpha \to \alpha \stackrel{\text{def}}{=} \forall(\alpha).(s \vee \alpha \to s \vee \alpha)$; upper bounds can be encoded, too, this time by intersections (e.g., $\forall (s \leq \alpha \leq t).\alpha \to \alpha \stackrel{\text{def}}{=} \forall(\alpha).((s \vee \alpha) \wedge t) \to (s \vee \alpha) \wedge t))$, but their frequency justifies the introduction of specific syntax.

The use of postfix **when** for variable quantification is borrowed from Typespec.

```
34  $ (integer() -> integer()) and
35    (true -> false) and (false -> true) and
36    (a -> a) when a: not(integer() or boolean())
```

and thus deduce for some function such as

```
37  def foo(x) when is_atom(x), do: negate(x)
```

the type `atom() -> atom()`, since an atom is neither an integer nor a Boolean.

It is possible to define polymorphic types with type parameters. For instance, we can define the type `tree(a)`, the type of nested lists whose elements are of type `a`, as

```
38  $ type tree(a) = (a and not list()) or [tree(a)]
```

and then use it to type the polymorphic function `flatten` that flattens a `tree(a)` returning a list of `a` elements:

```
39  $ tree(a) -> [a] when a: term()
40  def flatten([]), do: []
41  def flatten([x | xs]), do: flatten(x) ++ flatten(xs)
42  def flatten(x), do: [x]
```

The function above is well-typed. The three clauses of its definition are exhaustive (the last one captures all the arguments not captured by the first two). The first clause returns an empty list (thus a value of type `[a]`). The second clause captures any argument that is a non-empty list of `tree(a)` elements (since these are the only lists the function can be applied to), therefore our system deduces that `x` is of type `tree(a)` and `xs` is of type `[tree(a)]`; since `[tree(a)]` is a subtype of `tree(a)` (the latter being defined as the union of the former with another type), then both subsequent applications of `flatten` are well typed; therefore, both return results of type `[a]` whose concatenation (noted `++`) yields againt a result of type `[a]`. Finally, the inputs of type `tree(a)` that are not captured by the first two clauses, and are thus processed by the last clause, are just the arguments of type `a` that are not lists (from all inputs of type `tree(a)`, the first clause removes the empty lists, while the second clause removes all the non-empty lists of `tree(a)` elements); therefore the result of this clause is of type `[a and not list()]` and, by subtyping, of type `[a]`, too.

When `flatten` is applied, then the local type inference analyzes the argument, in order to determine the instantiation of the type of the function. If the argument is not a list, then `a` is instantiated to the type of the argument. If it is a list, then `a` is instantiated to the *union of the types of all the non-list elements of this nested list*. For instance, the type statically deduced for the application

```
43  flatten [3, "r", [4, [true, 5]], ["quo", [[false], "stop"]]]
```

is `[integer() or boolean() or binary()]` (where `binary()` is the type for strings).

**Protocols**   Elixir supports a kind of polymorphism akin to Haskell's typeclasses, via *protocols*. A protocol defines a set of operations that can be implemented for any type. For example, the **String.Chars** protocol requires the implementation of the `to_string` function. This function can convert any data type to a human representation as long as an implementation of the **String.Chars** protocol (viz., of `to_string`) has been defined for that data type. The *union* of all types that implement **String.Chars** is automatically filled in by the Elixir compiler and denoted by **String.Chars**.t(). In the absence of set-theoretic types this union would be approximated by `term()`.

Protocols can combine with parametric polymorphism to define more expressive types, such as collections. In Elixir, lists, sets, and ranges are all said to implement the **Enumerable** protocol which can be represented by the type **Enumerable**.t(a), that is, the enumerables whose elements are of type `a` (i.e., a-lists, a-sets, and a-ranges). So, for instance, the type of a generic `map` function that processes the elements of an enumerable will be type **Enumerable**.t(a), (a -> b) -> **Enumerable**.t(b), while the function that sums all elements of an enumerable of integer elements will have type **Enumerable**.t(integer()) -> integer().

The power behind Elixir protocols is that they allow library authors to express requirements in their APIs that are decoupled from the implementation of those requirements. For example, one strength of Elixir is in developing web applications. Web applications often have to encode Elixir data structures into different formats, such as JSON, CSV, XML, etc. To decouple the data types from the encoding logic, the author of a web framework defines a protocol, such as **JSON.Encoder**, and states it can encode any data structure, as long as it implements the **JSON.Encoder** protocol.

With set-theoretic types, library authors can now combine protocols to build additional requirements. One business may require that all of their public data must be available in several different data formats. They can encapsulate this requirement by defining an intersection of existing protocols:

```
44  $ type export() = JSON.Encoder.t() and CSV.Encoder.t() and XML.Encoder.t()
```

A system with looser requirements may use a union instead of intersection: the data type must implement at least one of the formats above, in order to be accepted by the system.

## 3  Extending Semantic Subtyping for Elixir

All features presented so far adapt to Elixir what is already possible in the type system of CDuce, defined via the set-theoretic interpretation of types of semantic subtyping [23]. There are however several key specific characteristics of Elixir that require the semantic subtyping framework to be modified, improved, and/or extended.

### 3.1  Function Arity

A first such characteristic is the arity of functions which plays an important role in Elixir. While it is possible to test the arity of a function using the expression `is_function` (e.g., `is_function(foo, 2)` tests whether `foo` is a binary function), it is not possible in semantic subtyping to express the type of exactly all functions with a specific arity.[7] This is because, in CDuce, all functions are unary, with a function that takes two arguments being considered a unary function that expects a pair. Although it is possible to define a type for all functions as `none() -> term()`,[8] it is not possible to give a type specifically for, say, binary functions using `{none(),none()} -> term()`, for the simple reason that a product of the empty set is equivalent to the empty set, and thus the latter type is equivalent to the former. To address this issue, we introduce a special syntax for function types, written as `(t₁,···,tₙ) -> t` which outlines the arity of the functions and that we already used in the previous examples. This allows the type of all binary functions to be written as `(none(),none()) -> term()`. However, this requires a non-trivial modification in the set-theoretic interpretation of function spaces: after defining the interpretation of multi-arity functions, subtyping is reframed as a set-containment problem. Solving this problem then produces the decision algorithm for subtyping (see Appendix A.1 for further details).

## 3.2 Guards and Pattern Matching

A second characteristic of Elixir that is not captured by the current research on semantic subtyping is the extensive use of guards both in function definitions and pattern matching.

In the previous examples, we have explicitly declared the type signature of all functions we defined, such as:

```
45  $ integer() -> integer()
46  def negate(x) when is_integer(x), do: -x
```

However, our type system is capable to infer the types of functions as the above even in the absence of their type declaration, by considering guards as explicit type annotations for the respective parameters. This not only applies to simple type tests of the parameters, but also to more complex tests. For example, for

```
47  def get_age(person) when is_integer(person.age), do: person.age
```

our system deduces from the guard that `person` must be a record with at least the field `age` defined, and containing a value of type `integer()`, that is, an expression of type `%{age: integer(), ...}`. This is a record type: records in Elixir are prefixed by `%` to distinguish them from tuples; the three dots indicate that the record type is open, that is, it types records where other fields may be defined (cf. Section 3.3). Since the dot in `person.age` denotes field selection, then our system deduces for the function `get_age` the type `%{age: integer(), ...} -> integer()`. One novelty of our system is that it can precisely express (most) guards in terms of types, in the sense that the set of values that satisfy a guard is the set of values that belong to a given type: for instance, the set of all values that satisfy the guard `is_integer(person.age)`) coincides with the set of values that have type `%{age: integer(), ...}`. Previous systems with semantic subtyping and set-theoretic types did not account for guards, which is the reason why we had to develop a specific analysis technique for them (see Section 4 for more details).

---

[7] In our system, to be able to express arity tests in terms of types is crucial for the precise typing of guards and, thus, of functions and pattern matching: cf. Section 3.2.

[8] The top type of functions of arity one is *not* `term()->term()`. In our system, every function of this type can be safely applied to any argument of type `term()`, that is, every well-typed argument. But of course not every function satisfies this property: only the total ones.

Note that, in the absence of such guards, it is the task of the programmer to explicitly provide the type of the whole function by preceding its definition by a type specification. It is also possible to elide parts of the return type of non-recursive functions by using the underscore symbol "_", as in

```
48  $ integer() -> _
49  def negate(x) when is_integer(x), do: -x
```

or

```
50  $ (integer() -> _) and (boolean() -> _)
51  def negate(x) when is_integer(x), do: -x
52  def negate(x) when is_boolean(x), do: not x
```

leaving to the type system the task of deducing the best possible types to replace for each occurrence of the underscore.

**Exhaustivity Checking**  Type analysis makes it possible to check whether clauses of a function definition, or patterns in a case expression, are *exhaustive*, that is, if they match every possible input value. For instance, consider the following code:

```
53  $ type result() =
54      %{output: :ok, socket: socket()} or
55      %{output: :error, message: :timeout or {:delay, integer()}}
56
57  $ result() -> string()
58  def handle(r) when r.output == :ok, do: "Msg received"
59  def handle(r) when r.message == :timeout, do: "Timeout"
```

We define the type `result()` as the union of two *record types*: the first maps the atom `:output` to the (atom) singleton type `:ok` and the atom `:socket` to the type `socket()`; the second maps `:output` to `:error` and `:message` to a union type formed by an atom and a tuple. Next consider the definition of `handle`: values of type `%{output: error, message: {:delay, integer()}}` are going to escape every pattern used by `handle`, triggering a type warning:

```
60  Type warning:
61    | def handle(r) do
62          ^^^^^^^^^
63       this function definition is not exhaustive.
64       there is no implementation for values of type:
65          %{output: :error, message: {:delay, integer()}}
```

Note that the type checker is able to compute the exact type whose implementation is missing, which enables fast refactoring since, as the type of `result()` or the implementation of `handle` are modified, the type checker will issue precise new warnings to point out the places where code changes are required.

**Redundancy Checking**    Similarly, it is possible to find useless branches—i.e., branches that cannot ever match. For instance, if we add a clause to the previous example:

```
66 $ result() -> string()
67 def handle(r) when r.output == :ok, do: "Msg received"
68 def handle(r) when r.message == :timeout, do: "Timeout"
69 def handle({:ok, msg}), do: msg
```

then since the specified input type is `result()` (which is a subtype of maps), the third branch will never match (its pattern matches only pairs) and can be deleted.

This will remove useless code, detect unused function definitions, or reveal more complex problems as these hints can indicate areas where the programmer's expectations and the actual logic of the program do not match.

**Narrowing**    Narrowing is the typing technique that consists in taking into account the result of a (type-related) test to refine (i.e., to narrow) the type of variables in the different branches of the test. In Section 2.2 we have already presented a simple example in which narrowing is used, namely, in the function `negate_alt` (code in line 17) the type-checker uses the test to narrow the type of `x`, which is (`integer()` `or` `boolean()`), to `integer()` in the "do" branch and to `boolean()` in the "else" branch. This is a simple application of narrowing, where the narrowing is performed on the type of a variable whose type is directly tested. However, our system is also able to narrow the type of the variables that occur in the expression tested by a "case" or a "if", even if this expression is not a single variable (some exceptions apply though: see future works). Here is a more complete example where we test the field selection on a variable

```
70 $ result() -> _
71 def handle(r) when r.output == :ok, do: {:accepted, r.socket}
72 def handle(r) when is_atom(r.message), do: r.message
73 def handle(r), do: {:retry, elem(r.message, 1)}
```

In the example the type of `r` which initially is `result()` is *narrowed* in the first branch to `%{output: :ok, socket: socket()}`, to `%{output: :error, message: :timeout}` in the second branch, and to `%{output: :error, message: {:delay, integer()}}}` in the last one. This precision is shown by the fact that `handle` type-checks the following type specification too:

```
74 $ (%{output: :ok, socket: socket()} -> {:accept, socket()}) and
75   (%{output: :error, message: :timeout} -> :timeout) and
76   (%{output: :error, message: {:delay, integer()}} -> {:retry, integer()})
```

As a matter of fact, deducing the type of the parameters of a function by examining its guards is just yet another application of narrowing where the function parameters are initially given the type `term()` and narrowed by the types deduced for the guards.

**Conservative Approximations**    When performing a type analysis on patterns with guards, it may not always be possible to determine the precise type of the captured values. In such cases, we use both lower and upper approximations to ensure that narrowing and exhaustivity/redundancy checking still work. As an example, consider the following simplistic function:

```
77  def foo(x) when map_size(x) == 2, do: Map.to_list(x)
```

We are unable to express by a type the exact domain of this function, which is the set of "all maps of size 2". However, when the guard succeeds, it is clear that x is a *map*, and this assumption is enough to deduce by narrowing that the body of the function is well-typed. Using the type of all maps to approximate the set of all maps of size 2 is an over-approximation. We call such a type the *potentially accepted type* of the pattern/guard since it contains all the values that *may* match it. Conversely, consider the following example:

```
78  def bar(x) when (is_map(x) and map_size(x) == 2) or is_list(x), do:
↪     to_string(x)
79  def bar(x) when length(x) == 2, do: x
```

The first clause matches both the maps of size 2 (but no other map) and any lists. Although we cannot characterize by a type all the values matched by the first clause, we do know that all lists are captured by it and, therefore, the second clause is redundant (`length` being defined only for lists). The type of all lists is an under-approximation (i.e., a subset) of the set of all values that satisfy the guard in the first clause. We refer to this under-approximation as the *surely accepted type* of the pattern/guard since it contains *only* values that *do* match it. Our system makes a distinction between guards that require approximation and those that do not, as further described in Section 4.

**Complex Guards**   The analysis of guards is more sophisticated than it appears. First of all, guards are examined left to right by incrementally generating environments during their analysis. An example is the guard in line 78: if we compare it with line 77 we see that we added an `is_map(x)` test. Without it the guard in line 78 would be equivalent to the one in line 77, since when x is a list, then `map_size(x) == 2` fails (rather than return `false`), and so does the whole guard: the `is_list(x)` would never be evaluated. To account for this, our analysis examines `is_list(x)` only if the preceding clause may not fail, which is always the case—though, it can return `false`— and `map_size(x)` is examined only in the environments in which `is_map(x)` succeeds.

Another stumbling block is that the analysis may need to generate for a single guard different type environments under which the continuation of the program is checked, as the following definition shows:

```
80  $ (term(),term()) -> {integer(),term()} or {term(),boolean()} or nil
81  def baz(x, y) when is_boolean(x) or is_integer(y), do: {y,x}
82  def baz(_, _), do: nil
```

The definition above type-checks, but this is possible only because the analysis of the guard `is_boolean(y) or is_integer(z)` in line 81 generates two distinct environments (i.e., one where z has type `integer()` and y type `term()`, and a second one where their types are inverted) which are both used to deduce *two* types for `{z,y}` which are then united in the result. By the same technique, in the absence of a type specification our system deduces for the first clause of `baz` in line 81 the type

```
83  ((term(),integer()) -> {integer(),term()}) and
84  ((boolean(),term()) -> {term(),boolean()})
```

and the analysis of the code defined in line 82 adds to this intersection the following arrow: `(not(boolean()),not(integer())) -> nil`.

Finally, our type system can analyze arbitrarily nested Boolean combinations of guards which are type tests of complex selections primitives, as the following definition of a parametric guard `is_data(d)` shows:

```
85  defguard is_data(d) when is_tuple(d) and tuple_size(d) == 2 and
86    (elem(d, 0) == :is_an_int and is_integer(elem(d, 1)) or
87     elem(d, 0) == :is_a_bool and is_boolean(elem(d, 1)))
```

Our system deduces that the guard `is_data(d)` succeeds if and only if `d` is of type `data()` defined as follows:

```
88  $ type data() = {:is_an_int, integer()} or {:is_a_bool, boolean()}
```

### 3.3 Records and Dictionaries

In Elixir, maps are a key-value data structure that serves as the primary means of storing data. There are two distinct use cases for maps: as records, where a fixed set of keys is defined, and as dictionaries, where keys are not known in advance and can be dynamically generated. A map type should unify both, allowing the type-checker to sensibly choose when it needs to ensure that some expected keys are present while enforcing type specifications for queried values.

**Maps as Records**  When used as records, Elixir provides the `map.key` syntax, where `:key` is an `atom()`. If the map returned by the expression `map` does not contain said key at runtime, an error is raised. In Section 3.2, we saw the following definition:

```
89  $ %{age: integer(), ...} -> integer()
90  def get_age(person) when is_integer(person.age), do: person.age
```

In more precise terms, the above type is equivalent to:

```
91  $ %{required(:age)=>integer(), optional(term())=>term()} -> integer()
```

Each "key type" in a map type is either required or optional. Singleton keys are assumed to be required, unless otherwise noted.[9] The triple dot notation means the type also types records values that define more keys than those specified in the type and corresponds to `optional(term()) => term()`. We refer to those as open maps. A program similar to the above but with optional keys

---

[9] The compiler will reject `required(term())`, as that would require a map with an infinite amount of keys. However, because a finite type such as `boolean()` can be either required or optional, we require all non-singleton types to be accordingly tagged to avoid ambiguity.

```
92  $ %{optional(:age) => integer()} -> _
93  def get_age(person), do: person.age
```

raises a type error pointing out the possibly undefined key:

```
94  Type warning:
95    | def get_age(person), do: person.age
96                                  ^^^^^^^^^^
97      key :age may be undefined in type: %{optional(:age) => integer()}
```

Hence typing gets rid of all `KeyError` exceptions for every use of *map.key*, by restricting the use of dot-selection to maps that are known to have the key.

**Maps as Dictionaries**   When working with maps as dictionaries, we use the `m[e]` syntax to access fields, where `m` and `e` are expressions returning a map and a key, respectively. In this notation, the field may not exist, in which case `nil` is returned. For the given function

```
98  $ %{optional(:age) => integer()} -> _
99  def get_age(person), do: person[:age]
```

the system infers `integer()` `or` `nil` as return type. If the type-checker can infer that the keys are present, then it will omit the `nil` return. For instance, the following function is well typed since both fields are required and, thus, cannot return `nil`:

```
100  $ %{foo: integer(), bar: integer()} -> integer()
101  def add(m), do: m[:foo] + m[:bar]
```

The type-checker distinguishes when a key *is* present, *may be* present, or *is not* present. To summarize, for the following function (where `%{}` is the empty map),

```
102  $ %{foo: integer()}, %{optional(:foo) => integer()}, %{} -> _
103  def f(m1, m2, m3), do: {m1[:foo], m2[:foo], m3[:foo]}
```

the return type `{integer(), integer() or nil, nil}` is inferred.

**The `fetch!` Operation**   Further interactions with maps happen via the `Map` module. We look at the `Map.fetch!(map, key)` function,[10] which expects an arbitrary key to exist in the map, raising a `KeyError` otherwise. A developer may use `fetch!` to denote that a given key was explicitly added in the past and it must exist at this given point. While `map.key` is ill-typed if the key *may be* undefined, `fetch!` is ill-typed only if the key *is always* undefined. So the following program is rejected

```
104  $ %{not_age: integer()} -> _
105  def get_age(m), do: fetch!(m, :age)
```

because the field `:age` cannot appear in `m`, but the following one is accepted

---

[10] In Elixir, ending a function name with `!` is a convention that implies the function may raise for valid domains. For example, `File.read!("log.txt")` will raise if the file does not exist, compared to `File.read("log.txt")` which would instead return `:error`.

```
106  $ map() -> term()
107  def get_age(m), do: fetch!(m, :age)
```

because key `:age` may be present in `m`, since `map()` represents any possible map.

**Dictionary Keys**   We have shown how to interact with maps as records and dictionaries where the keys were restricted to singleton types. But $e[e']$ (and other map operations) allows for generic use of maps as dictionaries, where the key is the result of an arbitrary expression $e'$. To model this behavior, map types can specify the type of values expected by querying over certain fixed key domains, e.g.,

```
108  $ integer() -> %{optional(integer()) => integer()}
109  def square_map(i), do: %{i => i ** 2}
```

The map `%{optional(integer()) => integer()}` associates integers to integers, but since `integer()` represents an infinite set of values, all fields cannot be required. Hence, it must be annotated as `optional`.

Key domains cannot overlap. To ensure this property in our system it is possible to use as key domains only a specific set of basic types of Elixir Typespec: `integer()`, `float()`, `atom()`, `tuple()`, `map()`, `list()`, `function()`, `pid()`, `port()`, and `reference()`. However, our type system allows the programmer to define map types that mix dictionary and record fields, that is, types where fields are declared both for singleton keys and for key domains. In that case it is possible to specify in the same type both some singleton keys and the domain keys of these singleton keys, the former taking precedence over the latter. This means that the type

```
110  $ type t() = %{ foo: atom(),
111                  optional(:bar) => atom(),
112                  optional(atom()) => integer() }
```

represents maps with a mandatory field `:foo` and an optional one `:bar` both set to atoms, and where any other key is an atom associated to an integer.

If $e$ is an expression of the type `t()` above, then the type of the selection $e[e']$ will be computed according to the type of the expression $e'$: if $e'$ has type `:foo`, then the selection has type `atom()`; if $e'$ has type `:bar`, then the selection will be typed by `atom() or nil`, since `:bar` may be absent; if the type of $e'$ intersects `atom()` but it is not a subtype of `:foo or :bar`, then the selection will be typed by `atom() or integer() or nil`. If $e'$ has type **not** `atom()` then the selection will have type `nil` and will issue a warning. If the `fetch!` operation is used in the examples above instead, then the system will deduce the same types minus `nil`, except for the case for $e'$ of type **not** `atom()`, which will be considered ill-typed. To summarize, if `m` is of type `t()`, then: `m.foo` has type `atom()`; `m.bar` is not well-typed; `m[:bar]` has type `atom() or nil`; `m[:other]` has type `integer() or nil`; `m[m.foo]` has type `atom() or integer() or nil`; `m[42+1]` has type `nil` and will raise a warning.

Finally, `t()` is a type that mixes the characteristics of records (i.e., the first two field declarations) and dictionaries (i.e., the last field declaration). This kind of mix is not uncommon in languages such as Lua*u* or TypeScript, where dictionaries sometimes include a mandatory field of some type (e.g., an integer field `size`). This is less common in Elixir and rather confined to Elixir's "structures" which are maps with a mandatory field `:__struct__` of type `atom()` and any other atom field optional, as in `struct %{__struct__: atom(), optional(atom()) => term()}`. A detailed comparison of our record types and those of Flow, TypeScript, or Lua*u* can be found in [8, Section 5].

**Deletions and Updates**    It is possible to specify missing keys in a map type by adding an optional field that points to `none()` (i.e., the key must be absent since if it were present, then it should be associated to a value of the empty type `none()`, which does not exist). This makes it possible to type a `delete` operation as follows:[11]

```
113  $ map() -> %{optional(:foo) => none(), ...}
114  def delete_foo(map), do: Map.delete(map, :foo)
```

Similarly to map access, there are different ways to replace a field in a map. The syntax `%{map | key => value}` requires that the key is present and otherwise raises a `KeyError` exception. The presence of such a key can be statically ensured by our system.

### 3.4 Gradual Typing and Strong Arrows

There is an important base of existing code for Elixir. If we want to migrate this code to a typed setting, the ability to blend statically typed and dynamically typed code is crucial. Some code that is working fine may not pass type-checking, therefore a gradual migration approach is preferred to converting the entire codebase to comply with static typing at once. This is the goal of gradual typing [47]. For that, we introduce the type `dynamic()`, which essentially puts the type-checker in dynamic typing mode. In practice, the programmer can think of `dynamic()` as a type that can become at run-time (technically, that *materializes* into: see [11]) *any other type*: an expression of type `dynamic()` can be used wherever any other type is expected, and an expression of any type can be used where a `dynamic()` type is expected since, in both cases, `dynamic()` may become at run-time that type.[12] The simplest use case is to declare that a parameter of a function has type `dynamic()`:

---

[11] The type of `delete_foo` is not very useful in practice. It will be useful when combined with "row polymorphism" which permits to check the following type: `%{ a } -> %{optional(:foo) => none(), a} when a : fields()` see future work.

[12] Oversimplifying, one can consider `dynamic()` to be both a supertype and subtype of every other type (while `term()`, which is often confused with `dynamic()` is only the former) with a caveat, subsumption does not apply to `dynamic()` since we cannot consider an expression of a type different from `dynamic()` to be of type `dynamic()`: the application of `dynamic()->dynamic()` to an integer is well-typed because the arrow type materializes in `integer()->dynamic()` and *not* because `integer()` materializes into `dynamic()`.

```
115 $ dynamic() -> _
116 def foo1(x), do: ⋯
```

meaning that in the body of `foo1` the parameter `x` can be given any type—possibly a different type for each occurrence of `x`—and that `foo1` can be applied to arguments of any type. The type `dynamic()` is a new basic type, that can occur in other type expressions. This can be used to constrain the types arguments can have. For instance

```
117 $ (dynamic() -> dynamic()) -> _
118 def foo2(f), do: ⋯
```

requires the argument of `foo2` to have a function type. This means that in the body of `foo2` the parameter `f` can be applied to arguments of any type and its result can be used in any possible context, but a use of `f` other than as a function—e.g., `f + 42`—will be rejected. Likewise, an application of `foo2` to an argument not having a functional type—e.g., `foo2({7,42})`—will be statically rejected, as well.

**Gradual Typing Guarantees** Using `dynamic()` does not mean that type-checking becomes useless. Even in the presence of `dynamic()` type annotations, our type system guarantees that if an expression is given type, say, `integer()`, then it will either diverge, or return an integer value, or fail on a run-time type-check verification. This safety guarantee characterizes the approach known as *sound gradual typing* [47]. This approach was developed for set-theoretic types in Lanvin's PhD thesis [30] whose results we use here to define subtyping and precision relations using the subtyping relation on non-gradual types (i.e., types in which `dynamic()` does not occur).

There is however a fundamental difference between our approach and the one of sound gradual typing: the latter uses the gradual type annotations present in the source code to *insert* into the compiled code the run-time type-checks necessary to ensure the above type safety guarantee. Instead, one of our requirements (cf. Section 5) is that the addition of types *must not* modify the compilation of Elixir. Therefore, we have to design our gradual type system so that it ensures the type safety guarantee by taking into account both the dynamic type-checks performed by the Erlang VM *and those inserted by the programmer*. The goal, of course, is to deduce for every well-typed expression a type which is gradual as little as possible, the best deduction being that of a non-gradual type (the less gradual the type, the more the errors captured at compile time). To that end we introduce the notion of *strong function types*.

**Strong Function Types** We have seen that our system can deduce the type of a function also in the absence of an explicit annotation when there are guards on the parameters. So from a typing point of view the following two definitions are equivalent since they both define the identity function of type `integer() -> integer()`:

```
119 $ integer() -> integer()        $ integer() -> integer()
120 def id_weak(x), do: x           def id_strong(x) when is_integer(x), do: x
```

However, from a runtime perspective, the two definitions above differ as the latter checks that its argument is of type `integer()`, while the former does not. Therefore, if these functions are applied to an argument that *may not be an integer* (e.g., of type `dynamic()`), then we can only be certain that the resulting output is an integer for the function `id_strong`. This distinction appears when we type the following function:

```
123 $ dynamic() -> {dynamic(), integer()}
124 def foo3(x), do: {id_weak(x), id_strong(x)}
```

which is accepted by our system since it deduces that when `id_weak` is applied to an argument of an unknown `dynamic()` type, then it cannot give to the result a type more precise than `dynamic()`, while for the same application with `id_strong` it deduces that whenever the application returns a result, this result will be of type integer.

We refer to the function `id_strong` as having a "strong" function type, since it guarantees that when applied to an argument that is *not within its domain*, it will either (*i*) return a result within its codomain, or (*ii*) fail on a dynamic type check—performed by the Erlang VM or inserted by the programmer—, or (*iii*) diverge.[13] Likewise, the function `id_weak` has a "weak" function type, since it may return (and actually, does return) a result not of type `integer()` when applied to an argument that is not of type `integer()`.

**Propagation of `dynamic()`**  The type `dynamic()` is propagated by the type system through the various calls of functions such as `id_weak`, but it is stopped when it goes through a function with a strong arrow type, such as `id_strong`: the type `dynamic()` does not appear in the type of the application of `id_strong`. Now, the goal of using `dynamic()` in a program is to instruct the type system to be lenient. In order to maximize this leniency, we propose (probably, as a type-checking option) to use an intersection to propagate `dynamic()` also for functions with a strong arrow type, in a way that increases the permissiveness of the type system without hindering its safety properties. However, this permissiveness comes at the expense of the static detection of some type errors.

In practice, this comes down to deduce for `foo3` a type more precise than the one given in line 123. If we omit the type annotation and, like for `foo3`, there is no explicit guard, then our system assumes that the parameter x has type `dynamic()` and deduces for `foo3` the type `dynamic() -> {dynamic(), (integer() and dynamic())}` which is a subtype of the type in line 123 (a classic sound gradual typing approach such as those by [47] or [11, 30] would have deduced for this function the return type `{integer(), integer()}`, but also modified its standard compilation by inserting two run-time integer type-checks, one for each occurrence of x in the body of `foo3`).

The reason why the intersection occurring in second projection of the result improves the typability of existing code can be generally understood by considering the type $t()$ **and** `dynamic()`. An expression of type $t()$ **and** `dynamic()` can be used not only in all contexts where an expression of type $t()$ is expected, but also in all contexts where a *strict* subtype of $t()$ is expected (in which case the use of `dynamic()` will be further propagated). This is useful especially for (strong) functions whose codomain is a union type. For instance, consider again the function `negate` as defined in lines 51–52. This is a strong function whose codomain is `integer()` **or** `boolean()`.

---

[13] If the argument is in the domain, then only (*i*) and (*iii*) are possible.

If this code is coming from some existing base—i.e., without any annotation—then the system deduces that this function takes a `dynamic()` input and returns a result of type (`integer()` **or** `boolean()`) **and** `dynamic()`: thanks to the intersection with `dynamic()` in the result type, it is then possible to use the result of `negate` not only where an expression of type `integer()` **or** `boolean()` is expected, but also where just an `integer()` or just a `boolean()` is expected, then propagating the dynamic type. Concretely, in this dynamic setting, the function `subtract` as defined in lines 4–6 would still be well typed since the type of `negate(b)` in line 5 could materialize into `integer()` and, in the absence of an explicit annotation, by the propagation of `dynamic()` the type deduced for the result of `subtract` would be `integer()` **and** `dynamic()` which thus in turn could be passed to a function expecting a subtype of `integer` (e.g., a `modulo` function expecting an input of type `integer()` **and not** `0`). Of course, this comes at the expenses of an earlier error detection, since a gradually typed version of `subtract` in which `negate` could be applied to a `boolean()` would fail: this is an error that would be instead statically detected in the absence of the propagation of `dynamic()`. Finally, notice that if we had explicitly defined the type of `negate` to be `dynamic() -> integer()` **or** `boolean()` (as we did above in line 123 for `foo3`), then the result of `negate(b)` in line 5 would have been typed as `integer()` **or** `boolean()` thus precluding the materialization and the consequent typing of `subtract`.

We extended the semantic subtyping framework with strong function types, which are inhabited by functions satisfying the property described above (see [9]). Strong function types are the key feature that allows the type-system to take into account the run-time typechecks, either performed by the VM or inserted by the programmer. Built-in operations, such as field selection, tuple projections, etc, are, by implementation, strong: the virtual machine dynamically checks that, say, if the field `a` of a value is selected, then the value is a record and the field `a` is defined in it. Our theory extends this kind of checks to user-defined operations—i.e., functions definitions—by analyzing their bodies to check that all the necessary dynamic checks are performed. The functions for which this holds have a strong type, and the system can safely deduce that when they are applied to an argument that *may* not be in their domain (e.g., an argument of type `dynamic()`), then the application will return a value in their codomain (rather than a result of type `dynamic()`), and as explained above, to maximize typability of existing code this codomain is intersected with `dynamic()`.

   All this is currently transparent to the programmer since strong types are only used internally by the type checker to deduce the type of functions such as `foo3`. A possible extension of our system would be to allow the programmer to specify whether higher-order parameters require a strong type or not.

## 4  A Pinch of Formalization

Elixir is, in essence, a minimalist language, with most of its constructs being syntactic sugar for the language's core expressions: functions and pattern matching. In this section we just outline the formalization of this core (in which a significant part of

| Base types | $b$ | $::=$ | $\texttt{int} \mid \texttt{atom} \mid \mathbb{1}_{\texttt{fun}} \mid \mathbb{1}_{\texttt{tup}}$ |
|---|---|---|---|
| Types | $t,s$ | $::=$ | $b \mid c \mid \alpha \mid \overline{t} \rightarrow t \mid \{\overline{t}\} \mid t \vee t \mid \neg t$ |
| Expressions | $e,f$ | $::=$ | $c \mid x \mid \lambda(\overline{x}.e) \mid f(\overline{e}) \mid \{\overline{e}\} \mid \texttt{elem}(e,e) \mid e+e$ |
| | | $\mid$ | $\texttt{let}\, x : t = e \,\texttt{in}\, e \mid \texttt{case}\, e \,\texttt{do}\, \overline{pg \rightarrow e}$ |
| Patterns | $p$ | $::=$ | $x \mid c \mid \{\overline{p}\}$ |
| Guards | $g$ | $::=$ | $g \,\texttt{and}\, g \mid g \,\texttt{or}\, g \mid \texttt{not}\, g \mid \texttt{is\_integer}(d)$ |
| | | $\mid$ | $\texttt{is\_atom}(d) \mid \texttt{is\_tuple}(d) \mid \texttt{is\_function}(d,d)$ |
| | | $\mid$ | $d == d \mid d\, != d \mid d < d \mid d <= d$ |
| Selectors | $d$ | $::=$ | $c \mid x \mid \texttt{elem}(d,d) \mid \texttt{tuple\_size}(d)$ |

■ **Figure 1** Expressions and Types

Elixir can be encoded), its typing and its extension to gradual typing. We omit the formalization of maps, the theoretical properties of the type system and its algorithmic aspects since they are fully detailed in two companion papers: [8] which formalizes the record and maps we presented in Section 3.3 and [9] which covers the aspects of function arity, guard analysis, and gradual typing.

### 4.1 Core Elixir

The syntax and types of Core Elixir are illustrated in Figure 1 where we use $c$ to range over constants (i.e., atoms or integers), $x$ to range over expression variables, $\alpha$ to range over type variables, and the notation $\overline{u}$ to denote the sequence $u_1, \ldots, u_n$.

Types are polymorphic, set-theoretic, and can be recursively defined (technically, they are the contractive regular terms coinductively generated by the grammar in Figure 1: see [7]). They are built from basic types (the types of all integers, all atoms, all functions, and all tuples, respectively), type variables, value constants (to represent singleton types), and by applying two constructors, for function types ($\overline{t} \rightarrow t$) and tuple types ($\{\overline{t}\}$) of given arity, and two connectives union and negation ($\vee$, $\neg$), with intersection $\wedge$ encoded as $t_1 \wedge t_2 = \neg(\neg t_1 \vee \neg t_2)$. We also encode the top type $\mathbb{1}$, the type of all values, as $\mathbb{1} = \texttt{int} \vee \texttt{atom} \vee \mathbb{1}_{\texttt{fun}} \vee \mathbb{1}_{\texttt{tup}}$, and the bottom type $\mathbb{0}$ as $\mathbb{0} = \neg\mathbb{1}$: they correspond to Elixir's `term()` and `none()` types, respectively.

Expressions include constants and variables, functions and applications, tuples and their projections, annotated let-expressions to model type annotations, and case expressions. The latter are composed by branches that are guarded by a pattern $p$ followed by a guard $g$. Patterns are either variables or constants or tuples thereof, while guards are Boolean combinations of tests of basic types and of relations on selector expressions.

The expressions of Core Elixir are translated into an intermediate calculus (defined in Figure 2 in Appendix B) in which all the negations in guards are eliminated (by pushing them to the leaves) and where all specific type tests are replaced by a generic one. This intermediate language can be typed by a type system (Figure 3 also in Appendix B) which is essentially a merge of the type system of polymorphic CDuce [15, 14] and of the type system for occurrence typing with set-theoretic types [12]. A sound algorithm to check whether an expression is well-typed in this type system is obtained simply by reusing the algorithmic techniques developed in the cited papers and embedding them in a bidirectional type system that uses the information of the type annotations in the let-expressions. There are however two exceptions which differentiate this system from the one in the cited works. First, in Elixir and, thus, in our core calculus, the index of a tuple projection can be the result of an expression, thus the typing rules for projections have to be reworked to take into account this aspect and our type system cannot statically ensure that the projection of a tuple will be in the bound of the tuple size (this happens only if any rule (proj$_\Omega$) in Figure 3 is used, in which case the compiler emits a warning). Second, and more importantly, is the use of guards, which are absent from previous work on semantic subtyping, and whose coverage requires new analysis techniques that we outline next.

## 4.2 Types for Patterns and Guards

Consider the expression `case e do` $(p_i g_i \rightarrow e_i)_{i<n}$. To type it we must type all the expressions $e_i$ of its branches. To give a precise type to each $e_i$ we must characterize the set of all values that match the pair $p_i g_i$, that is, $\{v \in \textbf{Values} \mid v \text{ matches } p_i g_i\}$. As discussed in Section 2.2, types denote sets of values; however, the converse does not hold, since not every set of values is exactly denoted by a type: the set of maps with exactly two fields cannot be expressed by a type. While the set of values matched by a pattern $p$ is a type, this is no longer true when patterns are combined with guards. We already met a combination of patterns and guards for which this does not hold in Section 3.2, where the first clause of the function `bar` in line 78 corresponds to using the following pattern-guard pair:

```
125   x when (is_map(x) and map_size(x) == 2) or is_list(x),
```

This pair matches all the maps of length two and all the lists but, as we know, there is no type that exactly denotes all maps of length two. In cases such as the above we approximate the set of all values that match a pattern-guard pair by *two* types: the smallest type larger than it, and the largest type smaller than it. In Section 3.2 we called these two types respectively the *potentially accepted type* and the *surely accepted type* of the guard. Here we introduce a notation for them, and note by $\l(pg\)$ and $\(pg\)$ the surely accepted type and the potentially accepted type of the pair $pg$. Using these two types we can compute the (best approximating) type $t_i$ that contains all values that *may* reach the expression $e_i$ in our initial case-expression. If the expression $e$ matched in the case-expression is of type $t$, then this type $t_i$ is defined as $(t \wedge \(p_i g_i\)) \setminus \bigvee_{j<i} \l(p_j g_j\)$. In words, the values in $t_i$ are those that may be produced by $e$ (i.e, those in $t$), and may be captured by $p_i g_i$ (i.e., those $\(p_i g_i\)$) and which are not surely captured by a preceding branch (i.e., minus those in $\l(p_j g_j\)$ for some $j < i$). This type $t_i$ can then be used to generate the type environment under which $e_i$ is typed: this environment, noted $t_i/_{p_i}$, assigns the type that can be deduced for each capture variable of the pattern $p_i$ under the hypothesis that the pattern is matched against a value in $t_i$.

As a first approximation, the typing of case-expressions can be summarized by the following typing rule:

$$(\text{case}_\Omega) \ \frac{\Gamma \vdash e : t \qquad (\forall i \leq n) \ (t_i \not\leq \mathbb{0} \ \Rightarrow \ \Gamma, \left(t_i/_{p_i}\right) \vdash e_i : s)}{\Gamma \vdash \text{case } e \text{ do } (p_i g_i \to e_i)_{i < n} : s} \ t \leq \bigvee_{i \leq n} \langle\!\langle p_i g_i \rangle\!\rangle$$

The rule types a case-expression with $n$ branches, where the $i$-th branch is guarded by the pattern $p_i$ and guard $g_i$. To type the expression $e_i$, the system computes $t_i$ as described above and produces the type environment $t_i/_{p_i}$. The latter is used to type $e_i$ only if $t_i \not\leq \mathbb{0}$, that is, only if the set of values that may be processed by the branch is not empty (this condition is used to check case redundancy). The rule has also the side condition $t \leq \bigvee_{i < n} \langle\!\langle p_i g_i \rangle\!\rangle$ which checks exhaustiveness: every possible result of $e$ (i.e., every value of type $t$) must be in the potentially accepted type of some branch. The name of the rule is marked by an $\Omega$, meaning that it will emit a warning: the union $\bigvee_{i \leq n} \langle\!\langle p_i g_i \rangle\!\rangle$ is an over-approximation of the set of values captured in the case-expression, so exhaustiveness may fail a run-time. If the stronger condition $\bigvee_{i \leq n} \langle\!\langle p_i g_i \rangle\!\rangle$ is also satisfied, then this cannot happen, and no warning is emitted (see the corresponding (case) rule in Figure 3 of Appendix B).

The typing rule for case expressions is actually more complicated than that, since it performs a finer-grained analysis of Elixir guards that is also used to compute their surely/potentially accepted types. Let us look at it in detail (where $\overset{L}{<}$ is the strict lexicographical order on pairs):

$$(\text{case}_\Omega) \ \frac{\Gamma \vdash e : t \qquad (\forall i \leq n) \ (\forall j \leq m_i) \ (t_{ij} \not\leq \mathbb{0} \ \Rightarrow \ \Gamma, \left(t_{ij}/_{p_i}\right) \vdash e_i : s)}{\Gamma \vdash \text{case } e \text{ do } (p_i g_i \to e_i)_{i < n} : s} \ t \leq \bigvee_{i \leq n} \langle\!\langle p_i g_i \rangle\!\rangle$$

$\textit{where } \Gamma ; t \vdash (p_i g_i)_{i \leq n} \rightsquigarrow (s_{ij}, \mathfrak{b}_{ij})_{i \leq n, j \leq m_i} \textit{ and } \ t_{ij} = (t \wedge s_{ij}) \setminus \bigvee_{\{(h,k) \,|\, (h,k) \overset{L}{<} (i,j) \text{ and } \mathfrak{b}_{hk}\}} s_{hk}$

The difference with the previous rule is that now system computes for each pair $p_i g_i$ a list of types $t_{i1}, ..., t_{im_i}$ rather than a single $t_i$. These $t_{i1}, ..., t_{im_i}$ are a partition of the previous $t_i$ and the rule types $e_i$ $m_i$-times, by generating each type environment $t_{ij}/_{p_i}$. The various $t_{ij}$'s are computed thanks to an auxiliary deduction system that computes them for all the branches of the case: $\Gamma ; t \vdash (p_i g_i)_{i \leq n} \rightsquigarrow (s_{ij}, \mathfrak{b}_{ij})_{i \leq n, j \leq m_i}$. This auxiliary judgment, whose definition is given in [9], essentially scans each $g_i$ from left to right looking for OR-clauses and, for each clause, it generates a list of pairs of the form $(s, \mathfrak{b})$ where $s$ is the type of the values for which the clause may be true and $\mathfrak{b}$ is a Boolean value that indicates whether $s$ is exact or not. For instance, for the guard (`is_map(x)` **and** `map_size(x) == 2`) **or** `is_list(x)` we used above, it will generate a list of two pairs: (`map()`, `false`) since the first clause may be true for some maps but not all of them, and (`list()`, `true`) since the second clause is true for all lists. Guards are parsed from left to right to take into account Elixir evaluation order and possible failures (a clause is typed only if the preceding clauses may not fail). The computation for a guard $g_i$ needs both $\Gamma$ and $p_i$ since $g_i$ can use variables that are in the environment or are introduced by $p_i$. Given $\Gamma ; t \vdash (p_i g_i)_{i \leq n} \rightsquigarrow (s_{ij}, \mathfrak{b}_{ij})_{i \leq n, j \leq m_i}$, then the potentially accepted type of $p_i g_i$ is the union of all $t_{ij}$'s, while the surely accepted type of $p_i g_i$ is the union of all $t_{ij}$'s for which $\mathfrak{b}_{ij}$ is true: $\langle\!\langle p_i g_i \rangle\!\rangle = \bigvee_{j \leq m_i} s_{ij}$ and $\langle\!\langle p_i g_i \rangle\!\rangle = \bigvee_{\{j \leq m_i \mid \mathfrak{b}_{ij}\}} s_{ij}$. On our example, if $g$ is the guard above, then $\langle\!\langle xg \rangle\!\rangle = \texttt{map()}$ **or** $\texttt{list()}$ and $\langle\!\langle xg \rangle\!\rangle = \texttt{list()}$, as expected. It is now clear how $t_{ij}$ is computed: it is the type of all values that are generated by $e$ and for which the $j$-th OR-clause of $g_i$ may evaluate to true (i.e., $t \wedge s_{ij}$) minus all the values that are surely captured either in a preceding branch or a preceding OR-clause of the branch (i.e., $\bigvee_{\{(h,k) \mid (h,k) \overset{L}{<} (i,j) \text{ and } \mathfrak{b}_{hk}\}} s_{hk}$).

## 4.3 Gradual Typing and Strong Arrows

We extend the previous system by adding gradual typing. This consists of adding to grammar for types in Figure 1 a new base type "?" (corresponding to the `dynamic()` type of Section 3.4) and extending the type system to account for strong arrows and the propagation of "?". This extension is not trivial and can be found in [9]. In what follows we briefly outline the main ideas of this extension.

To determine whether a function of type $s \to t$ is strong, noted $(s \to t)^*$, the rule $(\lambda^*)$ below checks whether the body $e$ of the function strongly ensures that it will return a result in $t$ (noted $e \,\texttt{::}\, t$) under the hypothesis that the parameter is *not* in the domain of the function (i.e., $x : \neg s$). This uses an auxiliary deduction system where, for instance, built-in operations such as the addition, typed by the rule (add) hereinbelow, are strong

$$(\lambda^*) \; \frac{\Gamma, x : s \vdash e : t \qquad \Gamma, x : \neg s \vdash e \,\texttt{::}\, t}{\Gamma \vdash \lambda(x.e) : (s \to t)^*} \qquad (\text{add}) \; \frac{\Gamma \vdash e_1 \,\texttt{::}\, \mathbb{1} \qquad \Gamma \vdash e_2 \,\texttt{::}\, \mathbb{1}}{\Gamma \vdash e_1 + e_2 \,\texttt{::}\, \texttt{int}}$$

and where the rule for case expressions does not check exhaustiveness (since if no branch matches, then the case fails and the expression is strong).

The addition of "?" is then handled by defining a *precision* relation $\preccurlyeq$. The intuition is that a type $t$ is more precise than a type $s$, written $s \preccurlyeq t$ if $t$ can be obtained by replacing in $s$ some occurrences of ? by other types.[14]

Whenever we need to use the precision relation to type an application, we propagate ?. If the function is weak, then the application can only be given type ?, while for strong functions we use the result type of the function intersected with ?, in order to improve the typability of gradually-typed programs:

$$\frac{\Gamma \vdash f : (s \to t)^* \quad \Gamma \vdash e : s' \quad s' \preccurlyeq s_1 \leq s_2 \succcurlyeq s}{\Gamma \vdash f(e) : \texttt{?} \wedge t} \qquad \frac{\Gamma \vdash f : s \to t \quad \Gamma \vdash e : s' \quad s' \preccurlyeq s_1 \leq s_2 \succcurlyeq s}{\Gamma \vdash f(e) : \texttt{?}}$$

In the future, we plan to experiment with different disciplines for ? propagation, for instance, to propagate ?, not only when the precision relation is needed, but also just when the types involved in the application have some ? components: this would enhance typability of untyped code but at the expense of the precision of the static detection of type errors and will have to be tested against existing bases of code.

---

[14] Actually, we use a semantic definition of $\preccurlyeq$, due to [30], which takes into account type equivalences: e.g., $(\{\texttt{?}, \texttt{int}\} \vee \{\mathbb{1}_{\texttt{fun}}, \texttt{?}\}) \preccurlyeq \{\mathbb{1}_{\texttt{fun}}, \texttt{int} \vee \texttt{atom}\}$ or $\neg\texttt{?} \preccurlyeq \texttt{?}$.

## 5   Integration into Elixir

**Requirements**   A type system tailored for Elixir needs to carefully balance the need to bring static error detection to existing users of the language without changing their experience, while still appealing to programmers coming from statically-typed languages. To meet these goals we established few basic requirements:

- the introduction of typing must not require any modification to the syntax of Elixir expressions (not even the addition of type annotations for function parameters);
- gradual typing should suffice to ensure that programmers are not forced to annotate existing code (apart from some corner cases);
- the system must extract a maximum of type information from patterns and guards; this should help finding type errors in current programs and encourage skeptical developers to provide types on more occasions;
- programmers who prefer a fully statically typed environment should be able to reduce (or remove altogether) the reliance on gradual types within their code by emitting warnings when `dynamic()` is used;
- we assume that most users who choose static typing over dynamic one would prefer to annotate all functions explicitly, for documentation and readability reasons; hence, priority is given to type most (if not all) Elixir idioms before turning our attention to more advanced (and computationally expensive) features such as type reconstruction.

Additionally, we set as an initial requirement that the addition of types must not modify the compilation of Elixir code. This guarantees the type system will not affect, for better or worse, the runtime behavior nor the performance of existing code. By clearly defining our scope, we can focus on the developer experience and ergonomics of the type system. In the future, we may lift this restriction and explore ways to use the type system to add runtime checks and drive performance improvements.

**Type Syntax**   Elixir is a language defined by minimal syntax with direct translation to Abstract Syntax Tree, similar to M-expressions introduced by McCarthy for LISP[35] A large part of Elixir is written in itself through macros, and therefore it does not provide special syntax for defining modules, functions, conditionals, etc.

From the typing perspective, this means Elixir shall not provide special syntax for types, and all the operators and notation found in types must match their uses outside of types, including associativity and precedence. While new operators can be introduced (as long as they consistently apply everywhere in the language), there is also a concern from the Elixir team about relying too much on punctuation and its impact on the language adoption.

With this in mind, we choose to use the operators **or**, **and**, and **not** to represent our fundamental set-theoretic operations. All of those operators exist in the language today and are extensively used in guards. Furthermore, we hope the Elixir community will find **JSON.Encoder**.t() **and** **XML.Encoder**.t() more readable than **JSON.Encoder**.t() & **XML.Encoder**.t(), and atom() **and** **not**(:foo **or** :bar) to be clearer than atom()\(:foo |:bar).

**Implementation**   We have implemented a prototype type-checker for Elixir, based on the formalization outlined in Section 4 and which is available at https://typex.fly.dev/. This prototype relies on a crucial component: a library of set-theoretic types, that checks the subtyping relation between types and solves type constraint problems. Our first prototype used the CDuce type library [17], but we are currently reimplementing it in Elixir itself, in order to deploy it within the Elixir compiler. A roadmap of the planned development of the type system of Elixir is given in the conclusion (Section 7).

## 6   Related Work

With this design, a typed Elixir would join the family of languages that make use of semantic subtyping, which includes CDuce [16] (for which it was originally designed), and newer additions such as Ballerina [2], Lu*au* [33, 28], and (partially) Julia [4].

The work most related to ours is the addition to set-theoretic types to Erlang by Schimpf et al. [45] which transposes and adapts the current polymorphic type system of CDuce to Erlang. Their work provides a nice formalization and a thorough comparison with other current type analysis and verification systems for Erlang. Essentially, [45] ends where we start from, namely the content of Section 2, whereas most of the features we presented in Section 3 are by [45] either left as future work (e.g., the typing of records and gradual typing) or ignored (e.g., the typing of function arity). A notable exception is the typing of patterns and guards: as in our case, and independently of us, the authors of [45] propose over and under approximations for types accepted by patterns combined with guards, but their analysis, which is similar to [26], stops to simple guards formed by conjunctions of type tests on single variables, thus avoiding the complex guard analysis we outlined in Section 4.2.

Another related work is eqWAlizer [21], an open-source Erlang type-checker (with an extremely succinct documentation) developed by Meta and used to check the code of WhatsApp. It consumes the spec and type alias of Erlang with few exceptions. In particular, and contrary to what we do, they have distinct types for records and dictionaries, and empty lists are subsumed to lists. As in our system, they use generics (constrained by the same **when** keyword we use) with local type inference, type narrowing, and gradual typing. In particular, eqWAlizer uses the same subtyping and precision relations for gradual types as we do, since both approaches are based on [30, 11]. However, eqWAlizer techniques to gradually type Erlang expressions are quite different from ours (no dynamic propagation or strong arrows). Another important difference is that when typing overloaded functions with overloaded specs (i.e., our intersections of arrow types) eqWAlizer does not take into account the order of the clauses of the functions while their applications require the argument to be compatible with a unique clause. Thanks to negation types our approach takes into account the order of clauses, and applications are correctly typed even if the argument is compatible with several clauses: it thus implements a more precise type inference.

Besides eqWAlizer and our approach, the theory of [11] is used also by Cassola et al. [6] to add gradual typing to Elixir. The work focuses on the gradual aspects, which is why the typing of the functional core is quite basic (no guards, no polymorphism, no set-theoretic types). For the gradual aspects, they use the sound gradual typing approach of [11] but they do not couple it neither with a cast-inserting compilation (to preserve Elixir semantics) nor with advanced techniques like ours that can take into account existing checks, and this may hinder the satisfaction of gradual guarantees.

The initial effort to type Erlang was by Marlow and Wadler [34] who defined a type system based on solving subtyping constraints. This type system supports disjoint unions, a limited form of complement, and recursive types, but not general unions, intersections, or negations, as we do. The formalization lacks proofs for first-class function types, which is a solved problem in semantic subtyping. One issue with this work is that they infer constrained types which are quite large, which leads to the use of a heuristics-based simplification algorithm to make them more readable.

Dialyzer [32], which serves as the current default to provide type inference in Erlang, is a static analysis tool that detects errors with a discipline of no false positive, while our static type system ensures soundness, that is, no false negative. Dialyzer lacks support for conditional types or intersection types to capture the relation between input and output types for functions, and record types are parsed but not used.

An actively developed alternative to type Erlang is Gradualizer [29], which also supports Elixir programs through a translation frontend. The approach looks similar to ours, though it lacks subtyping, with gradual typing inspired from [30]. But a comparison is difficult since it lacks a formalization or a detailed description.

The literature on Erlang also includes Hindley-Milner type systems [50] and bidirectional type systems (without set-theoretic types) [40].

Numerous statically-typed languages constructed for the Erlang VM have emerged over time. Two examples, Hamler and purerl [25, 38], derived from [39], incorporate a type system akin to Haskell's, including type classes. Notably, in Hamler, type classes are used to model OTP behaviors. Another language, Caramel [5], features a type system inspired from OCaml. Sesterl [46] extends the trend by offering a module system [43], utilizing functors to type OTP behaviors (a high priority in our future work list). Lastly, Gleam [24] is a functional language utilizing well-proven static typing methodologies from the ML community dating back to the early 90s: a Hindley-Milner type system [27, 36], supplemented with a rudimentary form of row polymorphism [51, 41].

## 7   Conclusion and Future Work

We presented the type system that we plan to incorporate in the Elixir compiler. The system is a transposition to languages of the Erlang family of the polymorphic type system of CDuce. To do that we had to improve and extend the latter to account for several characteristics of Elixir: the arity of functions, the use of guards, a uniform treatment of records and dictionaries, the need for a new sound gradual typing discipline that does not rely on the insertion at compile time of specific run-time type-tests but, rather, takes into account both the type tests performed by the virtual machine and those explicitly added by the programmer. The design of our system was guided by the principles and goals we briefly exposed in Section 5. Whether it achieves these goals will have to be checked on an actual implementation.

Incorporating a type system into a language used at scale can be a daunting task. Our concerns range from how the community will interact and use the type system to how it will perform on large code bases. Therefore, our plan is to introduce *very* gradually our gradual (pun intended) type system into the Elixir compiler.

In the first release types will be used just internally by the compiler. The type system will extract type information from patterns and guards to find the most obvious mistakes, such as typos in field names or type mismatches from attempting to add an integer to a string, without using any $-prefixed type specification: developers will not be allowed to write them. The goal is to assess the performance impact of the type system and the quality of the reports we can generate in case of typing violations, without tying the language to a specific type syntax.

The second milestone is to introduce type annotations only in *structs*, which are named and statically-defined closed record types. Elixir programs frequently pattern match on structs, which reveals information about the struct fields, but it knows nothing about their respective types. By propagating types from structs and their fields throughout the program, we will increase the type system's ability to find errors while further straining our type system implementation.

The third milestone is to introduce the $-prefixed type annotations for functions, with no or very limited type reconstruction: users can annotate their code with types, but any untyped parameter will be assumed to be of the `dynamic()` type.

The development of the type system will happen in parallel with further research into set-theoretic types and their application to other Elixir idioms, according to the lines we briefly describe next.

**Type Reconstruction and Occurrence Typing**   In the current system we can define and type JavaScript's "logical or" as follows:

```
126  $ ( (a , term()) -> a ) and ( (false or nil) , b -> b )
127    when a: not(false or nil), b: term()
128  def l_or(x, y) do: if x, do: x, else: y
```

The type is very precise: it states that when the first argument is of a type `a` that is neither `false` nor `nil` (as `not(false or nil)` is equivalent to `not false and not nil`), then the result is (of the type of) the first argument, otherwise it is (of the type of) the second argument. The type reflects Elixir's semantics of `if` which returns the `else:` part if and only if the tested value is either `false` or `nil`. By extending the techniques of [12] to polymorphic types, it will become possible not only to check but also to *reconstruct*—i.e., to infer in the absence of an explicit type specification—this same type for `l_or`. Using these same techniques we should be able to give more precise types to some common functions. For example, consider the classic `filter` function:

```
129  $ ((a -> boolean()) , [a]) -> [a] when a: term()
130  def filter(fun, []), do: []
131  def filter(fun, [h | t]) do
132    if fun.(h), do: [h | filter(fun, t)], else: filter(fun, t)
133  end
```

This function takes a predicate on the type `a`, a list of `a` elements, and returns the list of elements that satisfy the predicate. The definition above type-checks in our system, and the given type is as good a type as we can specify for it. However, by extending the techniques of [12] to polymorphic types, it will become possible to check (and probably also to reconstruct) for `filter` the following more precise type:

```
134  $ ((a and b -> true) and (a and not b -> false), [a]) -> [a and b]
135      when a: term(), b: term()
```

In our current system, checking this type fails because its verification requires to narrow the type of `h` in the branches of the if-expression by using the fact that, in the test, `h` is the argument of a function, `fun`, that has an intersection type; such a deduction requires the occurrence typing techniques developed by [12, 13] which can probably be decoupled from type reconstruction. Enabling type reconstruction in Elixir would make it easier for programmers to annotate their programs, since the type reconstructed for a function can be suggested as a starting point for the annotation (the type reconstructed for `l_or` would probably not be the first type a programmer would think of). Having a powerful type reconstruction system would also open the possibility for a strict type-checker mode that, instead of using `dynamic()` for non-annotated parameters, tries to infer their type. However, these advantages are counterbalanced by the computational price of this kind of reconstruction which makes several passes on the code each pass requiring the resolution of several type constraint problems. Therefore, a careful analysis of the costs and benefits of the approach must be performed before implementing it.

**Maps: Row Polymorphism and Key-types**   To write polymorphic annotations on functions operating on maps, row polymorphism is needed [51, 41], in particular when functions return results that are obtained by applying the field deletion and update operations on their inputs (see Footnote 11). However, extending semantic subtyping with row polymorphism is an open problem, and we are currently working on it.

Also, we plan to study how to remove the constraint that key-types must be chosen among a predefined set of types. Our idea is to allow the programmer to declare finite partitions of these predefined types and, eventually, to infer these partitions without an explicit declaration.

**Message-passing**   One key characteristic of Elixir is its concurrency and distribution system based on message-passing between lightweight threads called *processes*. Receiving messages from other processes is done through the **receive** construct, which relies on pattern matching and guards to match messages sitting in the process inbox. Typing the concurrency constructs and the actor model of Elixir is an obvious next step. Our type system is already capable of augmenting the code in **receive** with type information from guards, with narrowing and approximations. The *potentially accepted type* (cf. Section 3.2) of the patterns and guards in receive operations can be used to define *interfaces* (i.e., types) for processes and thus type higher-order communications. A longer-term research project is to type processes with behavioral types, in the sense of [1], for example by adapting the theory of Mailbox Types [19, 22].

**Behaviours**   Elixir is a language with first-class modules: modules can be passed as arguments to functions and returned as results. Modules are represented as atoms in the runtime and are currently dynamically typed as such. Furthermore, Elixir provides some sort of "static type" declarations for modules, called *behaviours* (British spelling). If you look at the documentation of any behaviour in the standard library of Elixir (cf. [20]), it consists of three parts: a list of types either concrete (i.e., the type is completely defined) or abstract (i.e., the type is just a name);[15] a list of *function callbacks*, that is function names alongside their domain and codomain which can use the types declared in the behaviour; a list of function definitions that are exported by the module and that cannot be modified. Borrowing the terminology from object-oriented programming, you can think of the callbacks as a list of abstract methods, and the function definitions as a list of final methods for a (abstract) class.

A module *implements* (i.e., it is typed by) a behaviour if it defines for each callback in the behaviour a function that accepts a superset of the callback domain and returns a subset of the callback codomain. Elixir uses behaviours to implement a naive static typing of modules: if a module adopting a behaviour does not implement all the (mandatory) callbacks of the behaviour or does not do it according to their specifications, then a warning is emitted. Currently, callbacks' domains and codomains are specified in Typespec's, but at some point (cf. the third milestone at the beginning of this section) we will want to use the types presented here for callbacks specifications, too. This will be the bare minimum: currently Elixir uses only the callback part of the information provided by behaviours and just to partially check new module implementations; but no check is performed when modules are passed around, not even that they are modules (just that they are atoms supposed to represent modules). Thus the next step will be to add behaviours as types of our system. This will allow us to reap some benefits of our static type system when programming with higher-order modules, by exploiting *all* information that behaviours provide.

This requires further research. As anticipated, behaviours may specify the type of callbacks in terms of abstract types, that is, nominal types whose concrete implementation is provided only by each module. In Elixir this corresponds to use `term()` in the Typespec specification of a callback. To fully exploit the information provided by behaviours, we need a type system capable of distinguishing the different roles of each type introduced in a behaviour: the concrete types, shared by all implementations, must be exported transparently; the abstract types, that can be processed and known only by the callback functions, must be exported opaquely; some other types, that are specific to each implementation but must be publicly known, must be instead parameters of the behaviours. A concrete example is given in Appendix C using mock-up syntax for the widely-used GenServer behaviour of Elixir standard library.

All the above requires extending the semantic subtyping framework, for instance to cope with existential types for packaged modules [44, 42] or Bounded Existential Types as implemented in Julia [52, 18]. At the same time, even a much simpler extension where abstract types were simulated by `dynamic()` will require careful consideration at the language design, especially in regard to backward compatibility.

---

[15] Actually, abstract types are not explicitly listed, but rather used in the specifications of the functions, by approximating them by `term()`.

**Acknowledgements** This work was partially supported by Supabase and Fresha. The second author was supported by a CIFRE grant agreement between CNRS and Remote Technology. The work benefited from the constant feedback from all the members of the Elixir compiler core development team.

## A Types

### A.1 Multi-argument Function Spaces

The definition of the multi-arity function spaces is defined as follows:

**Definition A.1.** Let $X_1, .., X_n$ and $Y$ be subsets of $D$.

$$(X_1, .., X_n) \to Y = \left\{ R : \mathscr{P}_f \left( D^n \times D_\Omega \right) \mid \forall (d_1, .., d_n, \delta) : R. \ (\forall i \in [1...n]. \ d_i : X_i \Longrightarrow \delta : Y) \right\}$$

With respect to the function space definition of semantic subtyping (cf., [23, Definition 4.2]), we replaced the single $D$ for the input with a $n$-ary product $D^n$, and enforced the constraint on the output $\delta$ only if *all* the elements of the input are in the expected domain.

Using the usual semantic subtyping relation for unary functions, our system encodes multi-arity function types using a CDuce record type:

**Encoding for Multi-arity Functions.**

$$\mathsf{Fun}((t_1, .., t_n), t) := \{ \mathsf{fn} = \{ t_1, .., t_n \} \to t; \ \mathsf{ar} = n \}$$
$$\mathsf{Fun}_{\mathbb{1}} := \{ \mathsf{fn} = \mathbb{0} \to \mathbb{1}; \ \mathsf{ar} = \mathtt{int} \}$$
$$\mathsf{Fun}_n := \{ \mathsf{fn} = \mathbb{0} \to \mathbb{1}; \mathsf{ar} = n \}$$

**Subtyping Properties.**
- $\forall (t_1, .., t_n, t) \quad \mathsf{Fun}((t_1, .., t_n), t) \leq \mathsf{Fun}_{\mathbb{1}}$
- there is a distinct top function for functions of all arity, which is for all $n$,

    $$\mathsf{Fun}_n = \{ \mathsf{fn} = \mathbb{0} \to \mathbb{1}; \mathsf{ar} = n \}$$

- since the integer singleton type $n$ is a subtype of $m$ if and only if $n = m$, each function type is discriminated by its arity:

    $$\mathsf{Fun}((t_1, .., t_n), t) \leq \mathsf{Fun}((s_1, .., s_m), s) \iff (n = m) \wedge (\forall i = 1 .. n, t_i \geq s_i) \wedge (t \leq s)$$

Note that this encoding makes intersections of function types of different arity be the empty type, since $n \wedge m$ is $\mathbb{0}$ if $n \neq m$.

## B  Non-gradual Language

### B.1  Syntax

$$i \text{ integers}, \quad k \text{ atoms}, \quad \alpha \text{ type variables}, \quad x \text{ variables}$$

| | | | |
|---|---|---|---|
| Constants | $c$ | $::=$ | $i \mid k$ |
| Expressions | $e, f$ | $::=$ | $c \mid x \mid \lambda(\overline{x}.e) \mid f(\overline{e}) \mid \{\overline{e}\} \mid \pi_e(e)$ |
| | | | $\mid \quad \mathtt{let}\, x : t = e \,\mathtt{in}\, e \mid e + e$ |
| | | | $\mid \quad \mathtt{case}\, e \,\mathtt{do}\, \overline{p\, g \to e}$ |
| Base types | $b$ | $::=$ | $\mathtt{int} \mid \mathtt{atom} \mid \mathbb{1}_{\mathtt{fun}} \mid \mathbb{1}_{\mathtt{tup}}$ |
| Types | $t, s$ | $::=$ | $b \mid c \mid \alpha \mid \overline{t} \to t \mid t \vee t \mid \neg t \mid \{\overline{t}\}$ |
| Singletons | $\ell$ | $::=$ | $c \mid \{\overline{\ell}\}$ |
| Patterns | $p$ | $::=$ | $\ell \mid x \mid \{\overline{p}\}$ |
| Guard atoms | $a$ | $::=$ | $\ell \mid x \mid \pi_a(a) \mid \mathtt{size}(a)$ |
| Guards | $g$ | $::=$ | $a\, ?\, t \mid a = a \mid a \mathrel{!=} a \mid g \,\mathtt{and}\, g \mid g \,\mathtt{or}\, g$ |
| Type environments | $\Gamma$ | $::=$ | $\bullet \mid \Gamma, x : t$ |

where no patterns have variables occurring more than once.

**■ Figure 2**  Expressions and Types

## B.2 Type System

$$\text{(cst)} \; \frac{}{c : c} \qquad \text{(var)} \; \frac{\dashv x : t}{x : t} \qquad \text{(+)} \; \frac{e_1 : \texttt{int} \quad e_2 : \texttt{int}}{e_1 + e_2 : \texttt{int}}$$

$$\text{($\lambda$)} \; \frac{\overline{x} : \overline{s} \vdash e : t}{\lambda(\overline{x}.e) : \overline{s} \to t} \qquad \text{(app)} \; \frac{f : \overline{s} \to t \quad \overline{e} : \overline{s}}{f(\overline{e}) : t}$$

$$\text{(proj)} \; \frac{f : i \quad e : \{t_0, .., t_n\}}{\pi_f(e) : t_i} \qquad \text{(proj$_\Omega$)} \; \frac{f : \texttt{int} \quad e : \{t_0, .., t_n\}}{\pi_f(e) : \bigvee_i t_i}$$

$$\text{(proj$_\Omega$)} \; \frac{f : \texttt{int} \quad e : \mathbb{1}_{\texttt{tup}}}{\pi_f(e) : \mathbb{1}} \qquad \text{(tuple)} \; \frac{\overline{e} : \overline{t}}{\{\overline{e}\} : \{\overline{t}\}} \qquad \text{(let)} \; \frac{f : s \quad x : s \vdash e : t}{\texttt{let } x : s = f \texttt{ in } e : t}$$

$$\text{(case)} \; \frac{\Gamma \vdash e : t \quad \forall i < n \; \forall j < m_i \text{ either } t_{ij} \leq \mathbb{0} \text{ or } \Gamma, \left(t_{ij}/p_i\right) \vdash e_i : s}{\Gamma \vdash \texttt{case } e \texttt{ do } (p_i g_i \to e_i)_{i<n} : s} \; t \leq \bigvee_{i<n} \langle\!\langle (pg)_i \rangle\!\rangle$$

$$\text{(case$_\Omega$)} \; \frac{\Gamma \vdash e : t \quad \forall i < n \; \forall j < m_i \text{ either } t_{ij} \leq \mathbb{0} \text{ or } \Gamma, \left(t_{ij}/p_i\right) \vdash e_i : s}{\Gamma \vdash \texttt{case } e \texttt{ do } (p_i g_i \to e_i)_{i<n} : s} \; t \leq \bigvee_{i<n} \langle\!\langle (pg)_i \rangle\!\rangle$$

*where* $\Gamma; t \vdash (p_i g_i)_{i \leq n} \rightsquigarrow (s_{ij}, b_{ij})_{i \leq n, j \leq m_i}$ *and* $t_{ij} = (t \wedge s_{ij}) \setminus \bigvee_{\{(h,k) \mid (h,k) \stackrel{L}{<} (i,j) \text{ and } b_{hk}\}} s_{hk}$

$$\text{(inst)} \; \frac{\Gamma \vdash e : t}{\Gamma \vdash e : t\varrho} \; \text{dom}(\varrho) \cap \Gamma = \varnothing \qquad \text{($\leq$)} \; \frac{e : t \quad t \leq s}{e : s} \qquad \text{($\wedge$)} \; \frac{e : t_1 \quad e : t_2}{e : t_1 \wedge t_2}$$

$$\text{($\vee$)} \; \frac{f : s \quad x : s \wedge u \vdash e : t \quad x : s \wedge \neg u \vdash e : t}{e[f/x] : t} \; \text{fv}(u) = \varnothing$$

■ **Figure 3** Declarative type system

The system of Figure 3 uses a presentation in which only the relevant part of the type environments is presented (i.e., the part $\Gamma \vdash$ is often omitted). In that system the rules marked by a "$\Omega$" correspond to cases in which the type-checker emits a warning since it cannot ensure type safety. More precisely, whenever a rule (proj$_\Omega$) is used the type-checker warns that the expression may generate an "index out of range" exception; when the rule (case$_\Omega$) is used the type-checker warns that the case may not be exhaustive.

 **The GenServer Behaviour**

GenServer is a behaviour of Elixir standard library that abstracts the common client-server interaction. It provides the boilerplate to supervise sync and async calls. The documentation of the latest version (v1.15.6) can be found at https://hexdocs.pm/elixir/1.15.6/GenServer.html.

A schematic description of its definition (written using Typespec) is given next, where we use colors to highlight the interesting parts

```
136  defmodule GenServer do
137    # Types
138    @type option() :: {:debug, debug()} | {:name, name()} | ...      //transparent
139    @type result() :: {:reply, reply(), state()} | ...              //transparent
140    @type state() :: term()                                         //opaque
141    @type request() :: term()                                       //parameter
142    @type reply() :: term()                                         //parameter
143      ⋮
144    # Callbacks
145    @callback init(init_arg :: term()) :: {:ok, state()} | :error
146    @optional_callback handle_call(request(),pid(),state()) :: result()
147    @optional_callback handle_cast(request(),state()) :: result()
148      ⋮
149    # Functions
150    @spec start(module(), any(), options()) :: on_start()           //higher-order
151      ⋮
152  end
```

Let us make some observations for each section:

- **Types.** The types `option()` and `result()` are completely defined and shared by all implementations of the GenServer behaviour, thus they must be transparently exported. The type `state()` is the type of internal state of a server and can be manipulated only by the functions of each implementation, thus it must be opaquely exported. The types `request()` and `reply()` describe the messages sent for requests and replies; as for `state()`, they are specific to each implementation, but they must be public so that processes can perform correct requests; therefore they are parameter of the implementations.

- **Callbacks.** Callbacks are either mandatory or optional. From a typing point of view this is akin to the optional and mandatory fields for map types.

- **Functions.** The first argument of the `start` function must be a module, but in practice it cannot be any module: it must be the same module that implements the behaviour or, at least (to type-check), a module implementing exactly the same behaviour.

What do these observations imply on the definition of types? If behaviours are to be promoted to (module) types, then we expect the GenServer behaviour to be defined as follows (we use a mock-up syntax combined with our types):

```
153  defmoduletype GenServer(request, reply) do
154    # Types
155    type option() = {:debug, debug()} | {:name, name()} | ...        //transparent
156    type result() = {:reply, reply, state()} | ...                    //transparent
157    type state()                                                      //opaque
158      ⋮
159    # Callbacks
160    callback init :: init_arg() -> {:ok, state()} | :error
161    callback optional(handle_call) :: request, pid(), state() -> result()
162    callback optional(handle_cast) :: request, state() -> result()
163      ⋮
164    # Functions
165    spec start :: GenServer(request, reply), init_arg(), options() -> on_start()
166      ⋮
167  end
```

The types `request` and `reply` are now parameters of the (module) type GenServer; the types `request()` and `reply()` are transparently exported, while `state()` is opaque. optional callbacks are explicitly declared with a syntax reminiscent of map types. Finally, the type of the first `start` function explicitly declares that the first argument of the function must be a module of the same type as the one implemented. Notice that the second argument of `start` is of type `init_arg()` since this argument is then passed to the function `init`. We left the definition of `init_arg()` unspecified, but it should probably be yet another parameter of the GenServer behaviour, exactly as `request()` and `reply()`.

### References

[1] Davide Ancona, Viviana Bono, Mario Bravetti, Joana Campos, Giuseppe Castagna, Pierre-Malo Deniélou, Simon J. Gay, Nils Gesbert, Elena Giachino, Raymond Hu, Einar Broch Johnsen, Francisco Martins, Viviana Mascardi, Fabrizio Montesi, Rumyana Neykova, Nicholas Ng, Luca Padovani, Vasco T. Vasconcelos, and Nobuko Yoshida. Behavioral Types in Programming Languages. *Foundations and Trends in Programming Languages*, 3:95–230, 2016. doi:10.1561/2500000031.

[2] Ballerina. https://ballerina.io/. Accessed on May 31, 2023.

[3] Véronique Benzaken, Giuseppe Castagna, and Alain Frisch. CDuce: an XML-centric general-purpose language. In *ICFP '03, 8th ACM International Conference on Functional Programming*, pages 51–63, Uppsala, Sweden, 2003. ACM Press. doi:10.1145/944705.944711.

[4] Jeff Bezanson, Jiahao Chen, Benjamin Chung, Stefan Karpinski, Viral B. Shah, Jan Vitek, and Lionel Zoubritzky. Julia: Dynamism and performance reconciled by design. *Proc. ACM Program. Lang.*, 2(OOPSLA), oct 2018. doi:10.1145/3276490.

[5] Caramel. https://caramel.run/. Accessed on May 31, 2023.

[6] Mauricio Cassola, Agustín Talagorria, Alberto Pardo, and Marcos Viera. A gradual type system for Elixir. In *Proceedings of the 24th Brazilian Symposium on Context-oriented Programming and Advanced Modularity*, pages 17–24, 2020. doi:10.1145/3427081.3427084.

[7] Giuseppe Castagna. Programming with union, intersection, and negation types. In Bertrand Meyer, editor, *The French School of Programming*. Springer, 2023. Preprint at arXiv:2111.03354.

[8] Giuseppe Castagna. Typing records, maps, and structs. *Proc. ACM Program. Lang.*, 7(ICFP), September 2023. doi:10.1145/3607838.

[9] Giuseppe Castagna and Guillaume Duboc. Guard analysis and safe erasure gradual typing: a type system for Elixir. Unpublished manuscript, December 2023.

[10] Giuseppe Castagna and Alain Frisch. A gentle introduction to semantic subtyping. In *Proceedings of the 7th ACM SIGPLAN international conference on Principles and practice of declarative programming*, pages 198–199, 2005. doi:10.1145/1069774.1069793.

[11] Giuseppe Castagna, Victor Lanvin, Tommaso Petrucciani, and Jeremy G Siek. Gradual typing: a new perspective. *Proceedings of the ACM on Programming Languages*, 3(POPL):1–32, 2019. doi:10.1145/3412932.3412940.

[12] Giuseppe Castagna, Mickaël Laurent, Kim Nguyen, and Matthew Lutze. On type-cases, union elimination, and occurrence typing. *Proceedings of the ACM on Programming Languages*, 6(POPL):75, 2022. doi:10.1145/3498674.

[13] Giuseppe Castagna, Mickaël Laurent, and Kim Nguyen. Polymorphic type inference for dynamic languages. Unpublished manuscript, January 2024.

[14] Giuseppe Castagna, Kim Nguyen, Zhiwu Xu, and Pietro Abate. Polymorphic functions with set-theoretic types. Part 2: local type inference and type reconstruction. In *Proceedings of the 42nd Annual ACM SIGPLAN-SIGACT Symposium on Principles of Programming Languages*, POPL'15, pages 289–302, January 2015. doi:10.1145/2676726.2676991.

[15] Giuseppe Castagna, Kim Nguyen, Zhiwu Xu, Hyeonseung Im, Sergueï Lenglet, and Luca Padovani. Polymorphic functions with set-theoretic types. Part 1: Syntax, semantics, and evaluation. In *Proceedings of the 41st Annual ACM SIGPLAN-SIGACT Symposium on Principles of Programming Languages*, POPL'14, pages 5–17, January 2014. doi:10.1145/2676726.2676991.

[16] CDuce. https://www.cduce.org/. Accessed on May 31, 2023.

[17] CDuce git repository. https://gitlab.math.univ-paris-diderot.fr/cduce/cduce. Accessed on May 31, 2023.

[18] Benjamin Chung, Francesco Zappa Nardelli, and Jan Vitek. Julia's efficient algorithm for subtyping unions and covariant tuples (pearl). In *33rd European Conference on Object-Oriented Programming, ECOOP 2019, July 15-19, 2019, London, United Kingdom*, volume 134 of *LIPIcs*, pages 24:1–24:15, 2019. doi:10.4230/LIPIcs.ECOOP.2019.24.

[19] Ugo de'Liguoro and Luca Padovani. Mailbox Types for Unordered Interactions. In *32nd European Conference on Object-Oriented Programming (ECOOP 2018)*, volume 109 of *Leibniz International Proceedings in Informatics (LIPIcs)*, pages 15:1–15:28, Dagstuhl, Germany, 2018. Schloss Dagstuhl–Leibniz-Zentrum fuer Informatik. doi:10.4230/LIPIcs.ECOOP.2018.15.

[20] Elixir. https://elixir-lang.org/. Accessed on May 31, 2023.

[21] eqWAlizer. https://github.com/WhatsApp/eqwalizer. Accessed on May 31, 2023.

[22] Simon Fowler, Duncan Paul Attard, Franciszek Sowul, Simon J. Gay, and Phil Trinder. Special delivery: Programming with mailbox types. *Proc. ACM Program. Lang.*, 7(ICFP), aug 2023. doi:10.1145/3607832.

[23] Alain Frisch, Giuseppe Castagna, and Véronique Benzaken. Semantic subtyping: dealing set-theoretically with function, union, intersection, and negation types. *Journal of the ACM*, 55(4):1–64, 2008. doi:10.1145/1391289.1391293.

[24] Gleam. https://gleam.run/. Accessed on May 31, 2023.

[25] Hamler. https://www.hamler-lang.org/. Accessed on May 31, 2023.

[26] Joseph Richard Harrison. *Robust Communications in Erlang*. PhD thesis, University of Kent, November 2020. doi:10.22024/UniKent/01.02.87484.

[27] J. Roger Hindley. The principal type-scheme of an object in combinatory logic. *Transactions of the American Mathematical Society*, 146:29–60, 1969. doi:10.2307/1995158.

[28] Alan Jeffrey. Semantic subtyping in Lua*u*. Blog post, November 2022. Accessed on May 6th 2023. URL: https://blog.roblox.com/2022/11/semantic-subtyping-luau.

[29] Svenningsson Josef. Gradualizer. https://github.com/josefs/Gradualizer. Accessed on May 31, 2023.

[30] Victor Lanvin. *A semantic foundation for gradual set-theoretic types*. PhD thesis, Université Paris Cité, 2021. URL: https://www.theses.fr/2021UNIP7159.

[31] Lukas Lazarek, Ben Greenman, Matthias Felleisen, and Christos Dimoulas. How to evaluate blame for gradual types. *Proc. ACM Program. Lang.*, 5(ICFP), aug 2021. doi:10.1145/3473573.

[32] Tobias Lindahl and Konstantinos Sagonas. Practical type inference based on success typings. In *ACM-SIGPLAN International Conference on Principles and Practice of Declarative Programming*, 2006. doi:10.1145/1140335.1140356.

[33] Lua*u*. https://luau-lang.org/. Accessed on May 31, 2023.

[34] Simon Marlow and Philip Wadler. A practical subtyping system for erlang. In *Proceedings of the Second ACM SIGPLAN International Conference on Functional Programming*, ICFP '97, page 136–149, New York, NY, USA, 1997. Association for Computing Machinery. doi:10.1145/258948.258962.

[35] John McCarthy. Recursive functions of symbolic expressions and their computation by machine, Part I. *Commun. ACM*, 3(4):184–195, apr 1960. doi:10.1145/367177.367199.

[36] Robin Milner. A theory of type polymorphism in programming. *Journal of Computer and System Sciences*, 17(3):348–375, 1978. doi:10.1016/0022-0000(78)90014-4.

[37] Sven-Olof Nyström. A soft-typing system for Erlang. In *Erlang Workshop*, 2003. doi:10.1145/940880.940888.

[38] Pure Erlang. https://github.com/purerl/purerl. Accessed on May 31, 2023.

[39] Purescript. https://www.purescript.org/. Accessed on May 31, 2023.

[40] Nithin Vadukkumchery Rajendrakumar and Annette Bieniusa. Bidirectional typing for Erlang. In *Proceedings of the 20th ACM SIGPLAN International Workshop on Erlang*, pages 54–63, 2021. doi:10.1145/3471871.3472966.

[41] Didier Rémy. Type checking records and variants in a natural extension of ML. In *Proceedings of the 16th ACM SIGPLAN-SIGACT Symposium on Principles of Programming Languages*, POPL '89, page 77–88, New York, NY, USA, 1989. Association for Computing Machinery. doi:10.1145/75277.75284.

[42] Andreas Rossberg. 1ML - core and modules united. *J. Funct. Program.*, 28:e22, 2018. doi:10.1017/S0956796818000205.

[43] Andreas Rossberg, Claudio Russo, and Derek Dreyer. F-ing modules. *Journal of functional programming*, 24(5):529–607, 2014. doi:10.1017/S0956796814000264.

[44] Claudio V. Russo. First-class structures for Standard ML. In *Proceedings of the 9th European Symposium on Programming Languages and Systems*, ESOP '00, page 336–350, Berlin, Heidelberg, 2000. Springer-Verlag. doi:10.1007/3-540-46425-5_22.

[45] Albert Schimpf, Stefan Wehr, and Annette Bieniusa. Set-theoretic types for erlang. In *Proceedings of the 34th Symposium on Implementation and Application of Functional Languages*, IFL '22, New York, NY, USA, 2023. Association for Computing Machinery. doi:10.1145/3587216.3587220.

[46] Sesterl. https://github.com/gfngfn/Sesterl. Accessed on May 31, 2023.

[47] Jeremy G. Siek and Walid Taha. Gradual typing for functional languages. In *Scheme and Functional Programming Workshop*, University of Chicago Technical Report TR-2006-06, pages 81–92, 2006.

[48] Erik Stenman. The Erlang runtime system. https://blog.stenmans.org/theBeamBook/. Accessed on May 31, 2023.

[49] Typespec. https://www.erlang.org/doc/reference_manual/typespec.html. Accessed on May 31, 2023.

[50] Nachiappan Valliappan and John Hughes. Typing the wild in Erlang. In *Proceedings of the 17th ACM SIGPLAN International Workshop on Erlang*, pages 49–60, 2018. doi:10.1145/3239332.3242766.

[51] Mitch Wand. Type inference for record concatenation and multiple inheritance. In *Proceedings of the Fourth Annual Symposium on Logic in Computer Science*, pages 92–97, 1989. doi:10.1109/LICS.1989.39162.

[52] Francesco Zappa Nardelli, Julia Belyakova, Artem Pelenitsyn, Benjamin Chung, Jeff Bezanson, and Jan Vitek. Julia subtyping: A rational reconstruction. *Proc. ACM Program. Lang.*, 2(OOPSLA), oct 2018. doi:10.1145/3276483.

## About the authors

**Giuseppe Castagna** is a CNRS Senior Research Scientist. His domain of expertise is the study of type-systems for functional programming languages.
https://orcid.org/0000-0003-0951-7535

**Guillaume Duboc** is a researcher at Remote Technology, responsible for designing, studying, and implementing a type system for the Elixir language, in the context of his PhD. studies.
https://orcid.org/0000-0002-6795-9545

**José Valim** is the creator of the Elixir programming language, an industrial project whose goal is to enable higher extensibility and productivity in the Erlang VM while keeping compatibility with Erlang's ecosystem.
https://orcid.org/0000-0001-8449-3997